\documentclass[aps,prb,twocolumn,superscriptaddress
]{revtex4-2}
\usepackage{chngcntr}
\usepackage{graphicx}
\usepackage{color}
\usepackage{amsmath}
\usepackage{enumitem}
\usepackage{amssymb}
\usepackage{hyperref}
\usepackage{cancel}
\usepackage{ulem}
\usepackage{multirow}
\usepackage{CJK}
\usepackage{verbatim}
\usepackage{dcolumn}
\usepackage{bm}
\usepackage{subfigure}
\usepackage{xcolor,float}

\definecolor{darkblue}{rgb}{0.,0.,0.4}
\definecolor{darkred}{rgb}{0.5,0.,0.}
\definecolor{BlueViolet}{RGB}{138,43,226}
\definecolor{SkyBlue}{RGB}{30,144,255}
\definecolor{DarkGreen}{RGB}{0,100,0}

\newcommand{\be}{\begin{equation}}
	\newcommand{\ee}{\end{equation}}

\newcommand{\bea}{\begin{eqnarray}}
	\newcommand{\eea}{\end{eqnarray}}

\begin{document}

\title{Microscopic study of 3D Potts phase transition via Fuzzy Sphere Regularization}

\author{Shuai Yang}	
\affiliation{Department of Physics and State Key Laboratory of Surface Physics, Fudan University, Shanghai 200433, P.R. China}
\author{Yan-Guang Yue}
\affiliation{Department of Physics and State Key Laboratory of Surface Physics, Fudan University, Shanghai 200433, P.R. China}
\author{Yin Tang}
\author{Chao Han}
\author{W. Zhu}
\email{zhuwei@westlake.edu.cn}
\affiliation{Institute of Natural Sciences, Westlake Institute for Advanced Study, Hangzhou 310024, China}
\affiliation{Department of Physics, School of Science, Westlake University, Hangzhou 310030, China }

\author{Yan Chen}
\email{yanchen99@fudan.edu.cn}
\affiliation{Department of Physics and State Key Laboratory of Surface Physics, Fudan University, Shanghai 200433, P.R. China}
\affiliation{Shanghai Branch, Hefei National Laboratory, Shanghai 201315, P.R. China}

\date{\today}

\begin{abstract}
The Potts model describes interacting spins with $Q$ different components, which is a direct generalization of the Ising model ($Q=2$). Compared to the existing exact solutions in 2D, the phase transitions and critical phenomena in the 3D Potts model have been less explored. 
Here, we systematically investigate a quantum $(2+1)$-D Potts model with $Q=3$ using a fuzzy sphere regularization scheme. We first construct a microscopic model capable of achieving a magnetic phase transition that separates a spin $S_3$ permutationally symmetric paramagnet and a spontaneous symmetry-breaking ferromagnet. 
Importantly, the energy spectrum at the phase transition point exhibits an approximately conformal symmetry, implying that an underlying conformal field theory may govern this transition. 
Moreover, when tuning along the phase transition line in the mapped phase diagram, we find that the dimension of the subleading $S_3$ singlet operator flows and drifts around the critical value $\sim 3$, which is believed to be crucial for understanding this phase transition, although determining its precise value remains challenging due to the limitations of our finite-size calculations. 
These findings suggest a discontinuous transition in the 3D 3-state Potts model, characterized by pseudo-critical behavior, which we argue results from a nearby multicritical or complex fixed point.   
\end{abstract}
	
\maketitle
	
\section{Introduction}
The enhanced symmetry usually occurs at the continuous phase transition point. A novel example is that, at the continuous phase transition point, the scale invariance increases to conformal symmetry \cite{polyakov1970conformal}. It leads to conformal field theory (CFT) as an effective low-energy theory governing the critical phenomena of second-order phase transitions \cite{Cardy_book,Henkel_book,yellowbook}.

In addition to continuous phase transitions, another class of transitions is first-order transitions, which are widespread in physical systems but have received much less attention. Since the correlation length remains finite, the system lacks exact scale invariance. It is generally believed that the first-order transition should not be described by some underlying scale-invariant effective theories. 
Recently, this traditional view has been revolutionized, i.e., it has been proposed that some of the weakly first-order transitions (``weak'' means a large but finite correlation length) belong to the pseudocritical regime governed by the complex CFT \cite{Gorbenko2018a,Gorbenko2018b,Kaplan2009} (compared to the real cases, complex CFT extends the definition of RG fixed points into the complex domain, with complex conformal data).
This idea has been successfully demonstrated in the 2D Potts model with $Q>4$ (details see Sec. II), where the emergent complex conformality and associated complex fixed points have been explicitly identified \cite{jacobsen2024lattice,tang2024}. 
In this context, the first-order phase transition in the original 2D Potts model (in the real parameter space), controlled by these complex fixed points, exhibits an approximate conformal symmetry inherited from the complex CFT. 
With this progress, a remaining puzzle is whether or not the above physical picture is applicable to the first-order transitions that happened in higher space-time dimensions, which motivates the current work.

Technically speaking, directly studying the higher-dimensional phase transition and evaluating critical exponents are challenging. Fortunately, the recently proposed fuzzy sphere regularization scheme offers a way out \cite{ZHHHH2022}. This scheme facilitates the extraction of various conformal data and has been successfully applied to the investigation of 3D Ising \cite{ZHHHH2022,OPE_hu2023}, Heisenberg \cite{han2023conformaloperatorcontentwilsonfisher}, and deconfined phase transitions \cite{zhou2024mathrmso5deconfinedphasetransition}.
In this work, by applying the fuzzy sphere technique, we propose a model that can realize spontaneous symmetry breaking of the $S_3$ symmetry, which should share the same universality class as the original 3D 3-state Potts model \cite{RevModPhys.54.235,Potts1952}. Employing Exact Diagonalization (ED) and Density Matrix Renormalization Group (DMRG) \cite{itensor} techniques, we are able to map out a global phase diagram and locate the phase boundary using the finite-size analysis of magnetic order parameters.  
Importantly, we extracted the operator spectrum along the transition line and observed an approximate conformal tower structure. 
It indicates that the 3D 3-state Potts phase transition is controlled by certain CFT. 
Moreover, despite of finite-size effect, we estimate the second scalar primary field $\epsilon'$ to be dangerously relevant. 
We deduce that the relevance of $\epsilon'$ leads to a sharp crossover to a pseudo-critical region due to the vicinity of a true fixed point just outside the model parameter space.

This paper is organized as follows: Sec. II briefly reviews the Potts model, focusing on the nature of transitions in general dimensions. In Sec. III, we introduce a microscopic model based on the fuzzy sphere to simulate the 3D 3-state Potts universality class. 
In Sec. IV, we determine the location and order of the phase transition based on the analysis of the order parameter. 
In Secs. V and VI, we provide the numerical evaluation of the operator spectrum, correlation functions, and operator product expansion coefficients. Sec. VII discusses the interpretation of the numerical results.

\section{Review of the Potts model}

The Potts model \cite{RevModPhys.54.235,Potts1952} describes local spins, living in space-time dimension $D$, oriented along $Q$ possible directions,  interact with their neighbors via short-range interactions.  
It is a generalization of the Ising model, which has wide applications in physics, such as the liquid-crystal, nematic-isotropic transition \cite{doi:10.1080/15421407108082773} and the structural cubic-to-tetragonal crystal transition \cite{PhysRevLett.37.565,WEGER19741}.

\begin{figure}[t]
    \centering
    \includegraphics[width=0.475\linewidth]{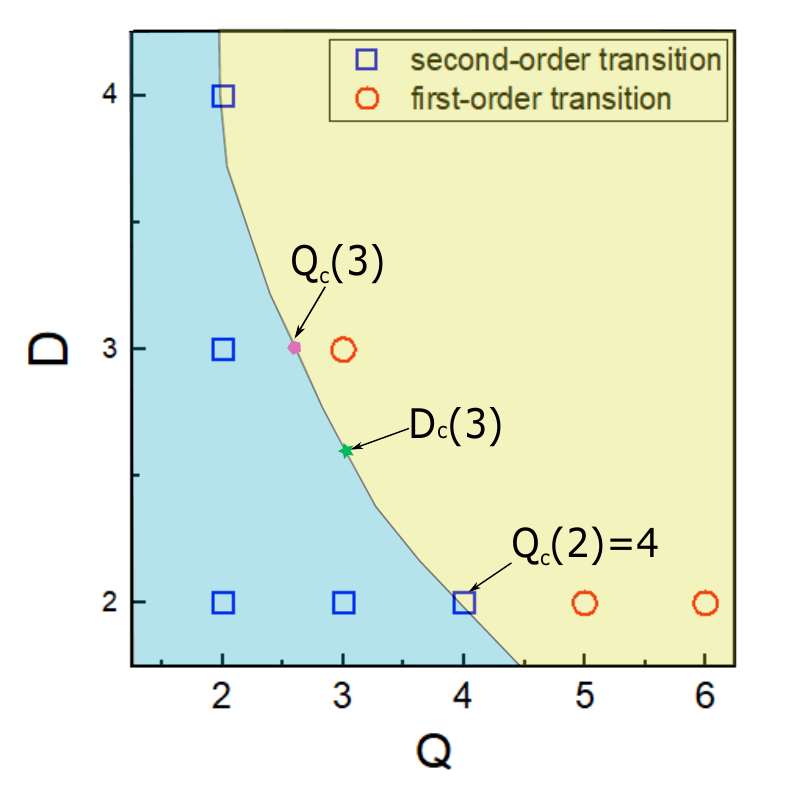}
    \qquad
    \begin{tabular}[b]{cc}\hline\hline
      $Q_c(3)$ &  Ref. \\ 
      2.7 &  $\epsilon$-exp \cite{RevModPhys.54.235} \\
      2.2 & RG \cite{Nienhuis1981} \\
      2.21 & MC \cite{Barkema1991} \\
      2.45 & MC \cite{Jooyoung1991} \\
      2.65 & MC \cite{Gliozzi2002} \\
      2.15 & SCOZA \cite{GROLLAU2001} \\
      2.11 & RG\cite{Villalobos2023} \\
      2.35 & MC \cite{Hartmann2005} \\
     \hline
      $D_c(3)$ &  Ref. \\ 
      2.6 &  CB \cite{chester2022} \\
     \hline \hline 
    \end{tabular}
    \caption{ (Left panel) The nature of Potts transition depends on space-time dimension $D$ and spin component $Q$ \cite{RevModPhys.54.235,Villalobos2023,Newman1984,Nienhuis1981,Hartmann2005,Barkema1991,Jooyoung1991,GROLLAU2001}, where the continuous transition exists for $Q\le Q_c, D\le D_c$. Here $Q_c, D_c$ are the largest critical values for which the Potts transition is continuous. 
    (Right panel) The estimated values of $Q_c(D=3)$ and $D_c(Q=3)$ from various literature. 
    } \label{fig:drawing}
\end{figure}

In 2D, the Potts model exhibits two families of CFTs with the same global symmetry, regarding the critical and tricritical branches, respectively. 
Early studies revealed that critical and tricritical fixed points can merge and annihilate with varying parameters $Q$ \cite{Cardy1980,Nauenberg1980,Pfeuty1981,Newman1984}. For the component \(Q\) exceeding a certain critical value \(Q_c(D)\) \cite{Cardy1980,Newman1984}, the phase transitions become the first-order type (see Fig. \ref{fig:drawing}). In particular, the critical value in 2D has been precisely determined to be $Q_c(D=2)=4$ \cite{Baxter1973,Nienhuis1980,Buddenoir1993}.   
Very recently, for the $Q>Q_c(2)$ Potts model, two fixed complex points have been identified by extending the physical parameter space to the complex plane \cite{Haldar2023,jacobsen2024lattice,tang2024}.  These complex fixed points control the first-order phase transitions in the original $Q>Q_c(2)$ Potts model in the real parameter space. Since the flow of the renormalization group between these complex fixed points is extremely slow, these weak first-order transitions appear almost indistinguishable from continuous ones \cite{Gorbenko2018a,Gorbenko2018b}. 
In this context, the first-order transitions in 2D Potts model are governed by complex CFTs accordingly. 

For the $D=3$ Potts model, the first-order phase transition is expected to appear for $Q>Q_c(3)$ as shown in Fig. \ref{fig:drawing} \cite{RevModPhys.54.235,Nienhuis1981,Villalobos2023,Newman1984}, but the critical value $Q_c(D=3)$ is not exactly known. Historically, the critical value $Q_c(3)$ 
has been estimated by various methods, such as $Q_c(3)\approx 2.7$ from the $\epsilon$-expansion \cite{RevModPhys.54.235}, $Q_c(3)\approx 2.2$ from the Kadanoff variational renormalization group \cite{Nienhuis1981}, $2<Q_c(3)<3$ from various Monte Carlo calculations \cite{Barkema1991,Jooyoung1991,Gliozzi2002} and $Q_c(3)\approx 2.11$ from the non-perturbative renormalization group \cite{Villalobos2023}.  
The Monte Carlo simulations partially support the above estimations \cite{3DPotts_MC_1979,3DPotts_MCRG_1979,Alves1991,JANKE1997,Berg2007} and the tensor renormalization group simulations \cite{Wang2014,NISHINO2000}, where the first-order transition is observed in the 3D $Q=3$-state Potts model. 
Moreover, based on recent numerical bootstrap study \cite{chester2022}, fixing $Q=3$, the critical space-time dimension separating the continuous and first-order transition is around $D_c\approx 2.6$. These results indicate that in the physical parameter space (see Fig. \ref{fig:drawing}), the 3D 3-state Potts model ($D=3,Q=3$) is sitting close to the phase boundary.  Additionally, the conformal fixed points merge-and-annihilate picture is also supported by the calculation based on the non-perturbative renormalization group in higher dimensions \cite{Villalobos2023,Newman1984,Nienhuis1981}.  
Inspired by these facts, a natural conjecture is a first-order transition in the 3D $ 3$ state Potts model can also be described by the complex CFT, akin to the 2D $(Q>4)$-state Potts model.

It is worth noting that the existing results for $D=3$ case plotted in Fig. \ref{fig:drawing} mainly come from numerical simulations on a particular model (lattice Potts model with nearest neighbor couplings), or perturbative calculations. The conclusion with $D =3 ,Q = 3$ that undergoes a first-order transition \cite{3DPotts_MC_1979,3DPotts_MCRG_1979,Alves1991,JANKE1997,Berg2007,gaite2024} is not proved, or it is still an open question if or not all lattice models undergo a first-order phase transition (see Sec. VII). 
So it is desired to construct other models realizing 3D Potts universality class, and to construct a global phase diagram with more tuning parameters \cite{Gavai1992}.  We believe inspecting the nature of 3D Potts transition from a different angle would be helpful.

\begin{figure}[b]
	\includegraphics[width=0.47\textwidth]{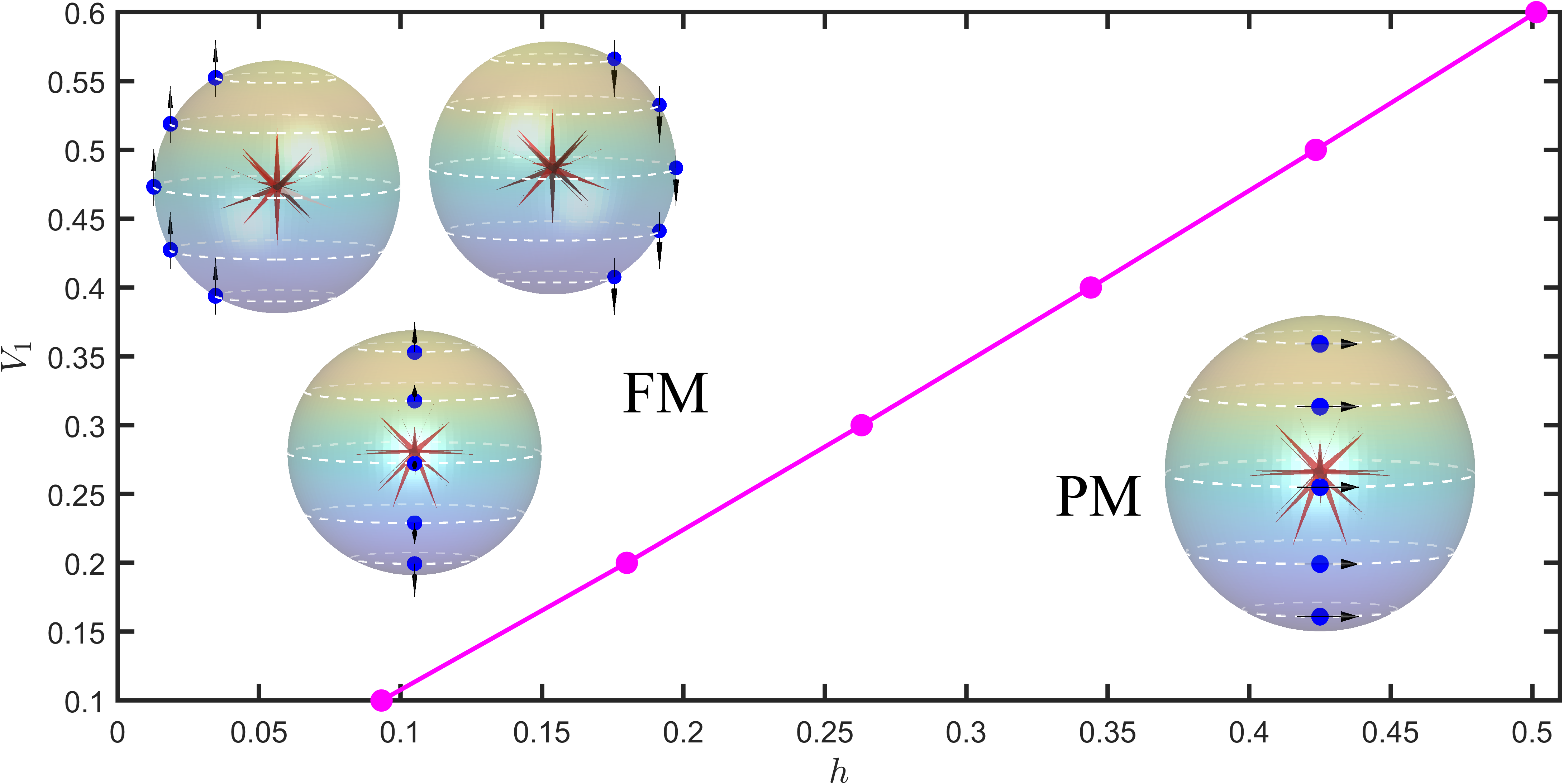}
	\caption{ A schematic plot of phase diagram with a phase transition separating a paramagnet ($h>h_c$) from a symmetry-breaking ferromagnet ($h<h_c$). The phase transition points are determined as: $V_1=0.1:h_c\approx0.0933;V_1=0.2:h_c\approx0.18;V_1=0.3:h_c\approx0.263;V_1=0.4:h_c\approx0.344;V_1=0.5:h_c\approx0.4235;V_1=0.6:h_c\approx0.5015$.
  }
	\label{fig:phasediagram}
\end{figure}

\section{3-State Potts model on fuzzy sphere}
Drawing upon the previous Ising model \cite{ZHHHH2022} with a global $Z_2$ symmetry, we consider interacting fermions with three flavors to achieve the full $S_3$ permutation symmetry. These fermions live on the fuzzy sphere, with a magnetic monopole of charge $4\pi s$ placed at the center. The system is characterized by a continuous Hamiltonian:\
\begin{widetext}
\begin{equation}\label{eq:ham}
H=\int{d\Omega_ad\Omega_bU\left( \Omega_{ab} \right)\left[ n_0\left( \Omega_a \right) n_0\left( \Omega_b \right) -n_z\left( \Omega_a \right) n_{z}^{\dag}\left( \Omega_b \right) \right]}
-h\int{d\Omega \left[ n_x\left( \Omega \right) +n_{x}^{\dag}\left( \Omega \right) \right]}
\end{equation}
where $\Omega=(\theta,\psi)$ is spatial coordinates on a sphere with radius $R$, density operator reads $n_\alpha (\Omega)=\psi^{\dagger}(\Omega)S_{\alpha}\psi(\Omega)$, and $S_{\alpha}$ is defined in $Q=3$ local Hilbert space
\begin{equation}
S_0=\left(\begin{array}{ccc}
1 & 0 & 0 \\
0 & 1 & 0 \\
0 & 0 & 1
\end{array}\right),
S_z=\left(\begin{array}{ccc}
1 & 0 & 0 \\
0 & e^{i \frac{2 \pi}{3}} & 0 \\
0 & 0 & e^{i \frac{4 \pi}{3}}
\end{array}\right), 
S_x=\left(\begin{array}{ccc}
0 & 0 & 1 \\
1 & 0 & 0 \\
0 & 1 & 0
\end{array}\right).
\end{equation}
The interaction term $U\left(\Omega_{a b}\right)=\frac{g_0}{R^2} \delta\left(\Omega_{a b}\right)+\frac{g_1}{R^4} \nabla^2 \delta\left(\Omega_{a b}\right)$ is taken to be local and short-ranged, ensuring that the phase transition is described by a local theory. After projecting to the lowest Landau level (LLL), the second quantized Hamiltonian can be derived in the following form:
\begin{equation}
H=\sum_{m_1m_2m_3m_4}{V_{m_1m_2m_3m_4}\left[\left( \boldsymbol{c}_{m_1}^{\dag}\boldsymbol{c}_{m_4} \right) \left( \boldsymbol{c}_{m_2}^{\dag}\boldsymbol{c}_{m_3} \right)-\left( \boldsymbol{c}_{m_1}^{\dag}S_z\boldsymbol{c}_{m_4} \right) \left( \boldsymbol{c}_{m_2}^{\dag}S_z^{\dag}\boldsymbol{c}_{m_3} \right)\right]-h\sum_m{\boldsymbol{c}_{m}^{\dag}\left( S_x+S_x^{\dag} \right) \boldsymbol{c}_m}}
\end{equation}
where $\boldsymbol{c}_m^{\dagger}=\left(c_{m 0}^{\dagger}, c_{m 1}^{\dagger}, c_{m 2}^{\dagger}\right)$ is the fermion creation operator on the $m_{\text{th}}$ Landau orbital, and $V_{m_1,m_2,m_3,m_4}$ is associated with Haldane pseudopotential $V_l$ with following form:
\begin{equation}
V_{{m_1m_2m_3m_4}}=\sum_lV_l(4s-2l+1)\begin{pmatrix}
s&s&2s-l\\m_1&m_2&-m_1-m_2
\end{pmatrix}\begin{pmatrix}
s&s&2s-l\\m_4&m_3&-m_4-m_3
\end{pmatrix}\delta_{m_1+m_2,m_3+m_4},
\end{equation}
where $\left(\begin{array}{ccc}j_1 & j_2 & j_3 \\ m_1 & m_2 & m_3\end{array}\right)$ is the Wigner 3j symbol. In this paper, we will only consider ultra-local interactions $U$ in real space, which corresponds to
non-zero Haldane pseudopotentials $V_0$ and $V_1$.

\end{widetext}

This Hamiltonian possesses complete global $S_3$ permutation symmetry, which can be decomposed into a cyclic operation, i.e., $Z_3$ symmetry that remains invariant under cyclic permutation of the three local spin degrees of freedom--$\left|0\right>,\left|1\right>,\left|2\right>$, and a unitary charge conjugation operation, i.e., $Z_2$ symmetry that is invariant under the exchange of two spin states $\left|1\right>$ and $\left|2\right>$. This Hamiltonian also respects $SO(3)$ rotational symmetry in real space.

The number of Landau orbitals of the lowest Landau level is equal to $N_{\mathrm{o}}=2s+1$, and we will consider the number of filled electrons as $N_{\mathrm{e}}=2s+1$. 
For systems with orbital numbers \( N_{\mathrm{o}} = 4-6 \), we use ED to obtain the entire energy spectrum. For systems with \( N_{\mathrm{o}} = 7 \) and \( N_{\mathrm{o}} = 8 \), we use ED to solve for the first 4000 eigenstates. For systems with \( N_{\mathrm{o}} = 9-14 \), we use DMRG to obtain the first 30 eigenstates. During the DMRG calculations, the maximum bond dimension is set to \( \chi = 2500 \).
For systems sizes with \( N_o = 9\sim 12 \), once the energy error of the eigenstates obtained from 10 consecutive sweeps is less than \( 10^{-10} \), and for larger systems with \( N_o = 13\sim 16 \), the maximum tolerance is set to \( 10^{-8} \), we consider it a faithful representation of the accurate eigenstates.

At this filling ($N_{\mathrm{o}}=N_{\mathrm{e}}$), the interlayer interaction among the electrons selects a ferromagnetic state as the ground state, i.e. electrons tend to occupy the same flavor color with two others empty. 
In contrast, the transverse field term $h$ leads to a quantum paramagnet as the ground state, i.e., electrons equally occupy the three flavors; thus, no net magnetization is expected. Combining the density interaction and transverse field term in Eq. \eqref{eq:ham}, a direct transition between ferromagnet and paramagnet is expected, as shown in Fig. \ref{fig:phasediagram}. 
A typical feature of the phase diagram is a phase transition line (instead of a single transition point) separating ferromagnet and paramagnet. This is achieved by adding two interaction terms regarding parameters $V_0,V_1$. Next, we will study the nature of this transition line in Sec. V, which offers more information on this problem. As a comparison, the existing work in the lattice Potts model with nearest neighbor couplings \cite{3DPotts_MC_1979,3DPotts_MCRG_1979,Alves1991,JANKE1997,Berg2007} contains only a single transition point.

\begin{figure}[b]
\includegraphics[width=0.49\linewidth]{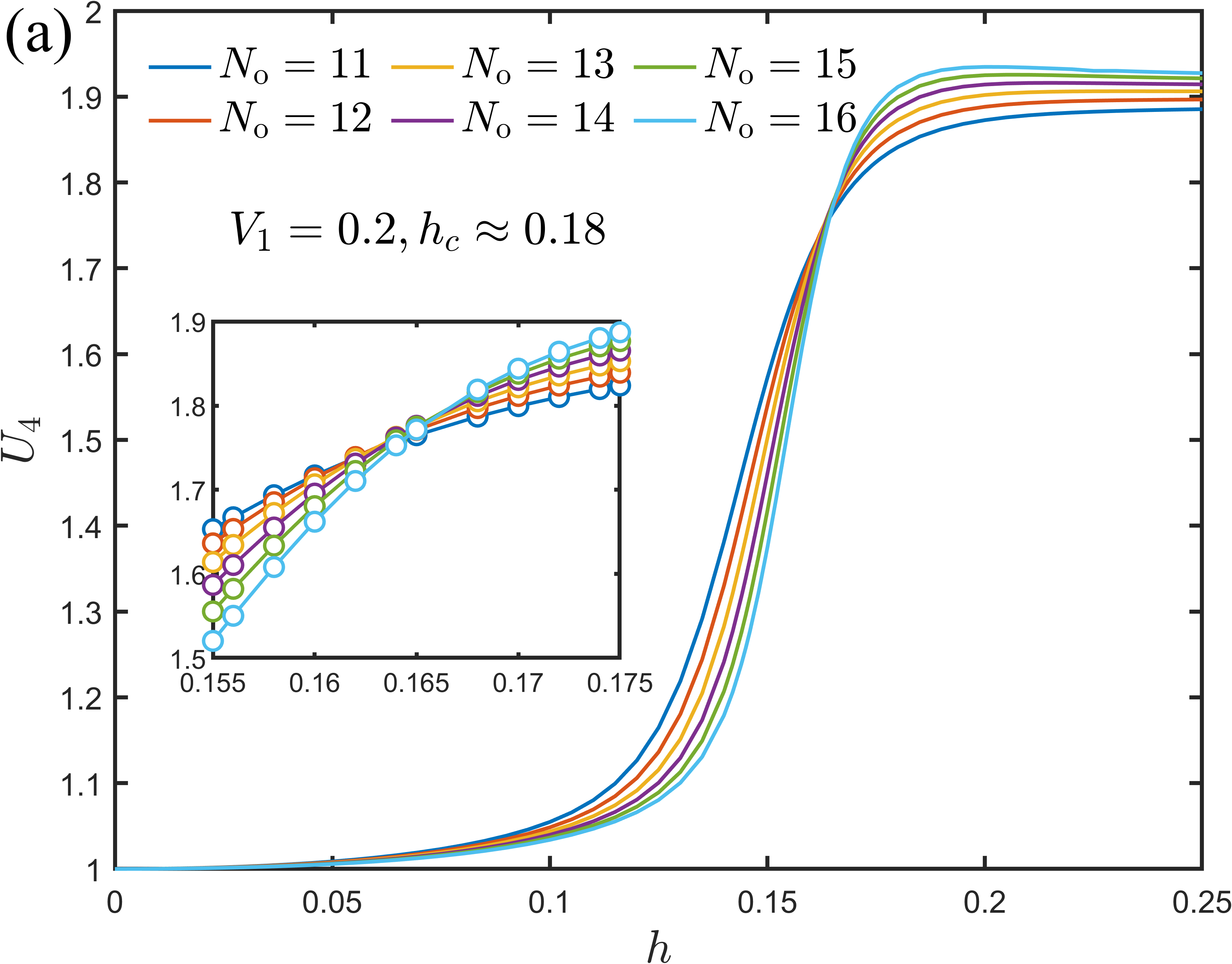}
\includegraphics[width=0.49\linewidth]{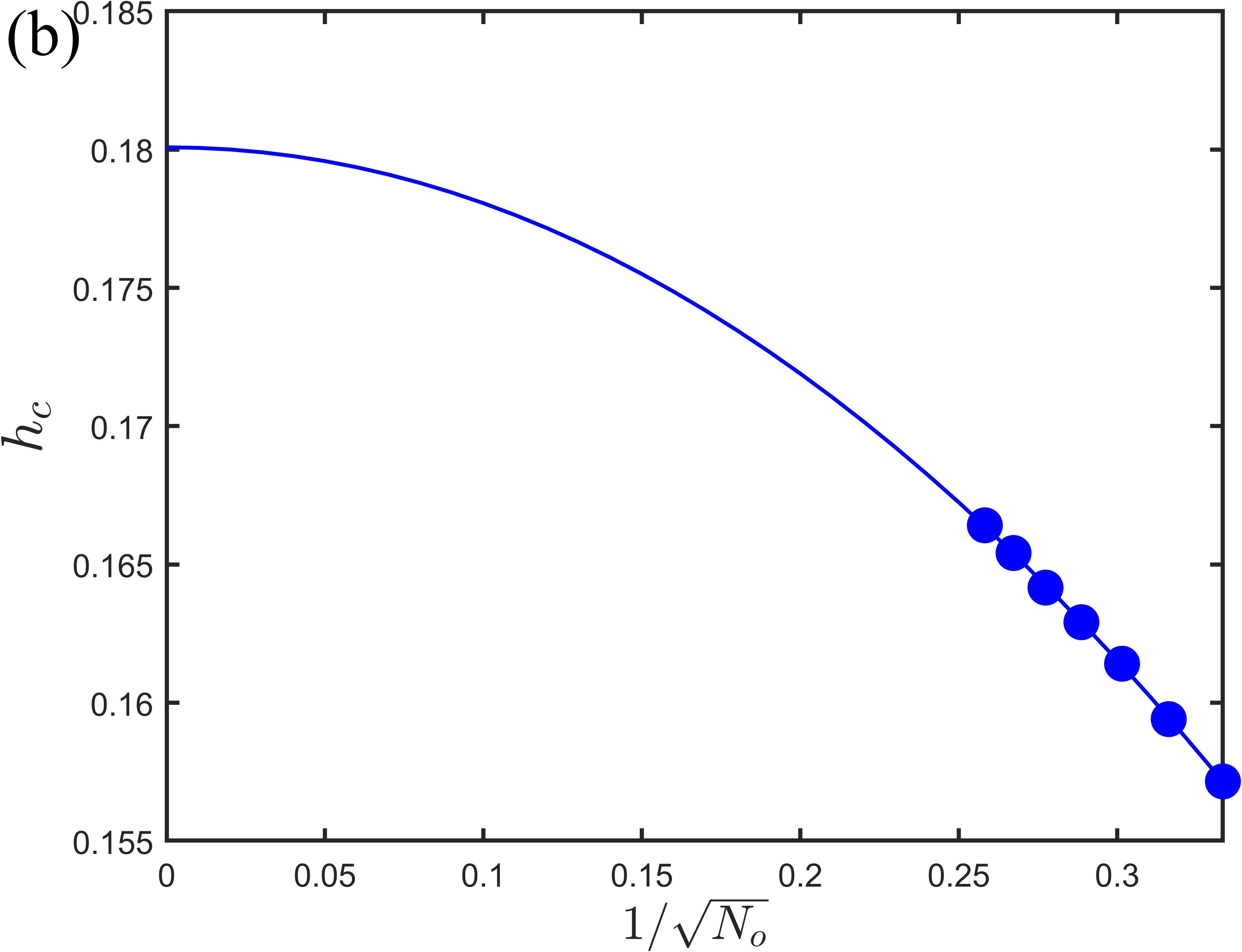}
\includegraphics[width=0.49\linewidth]{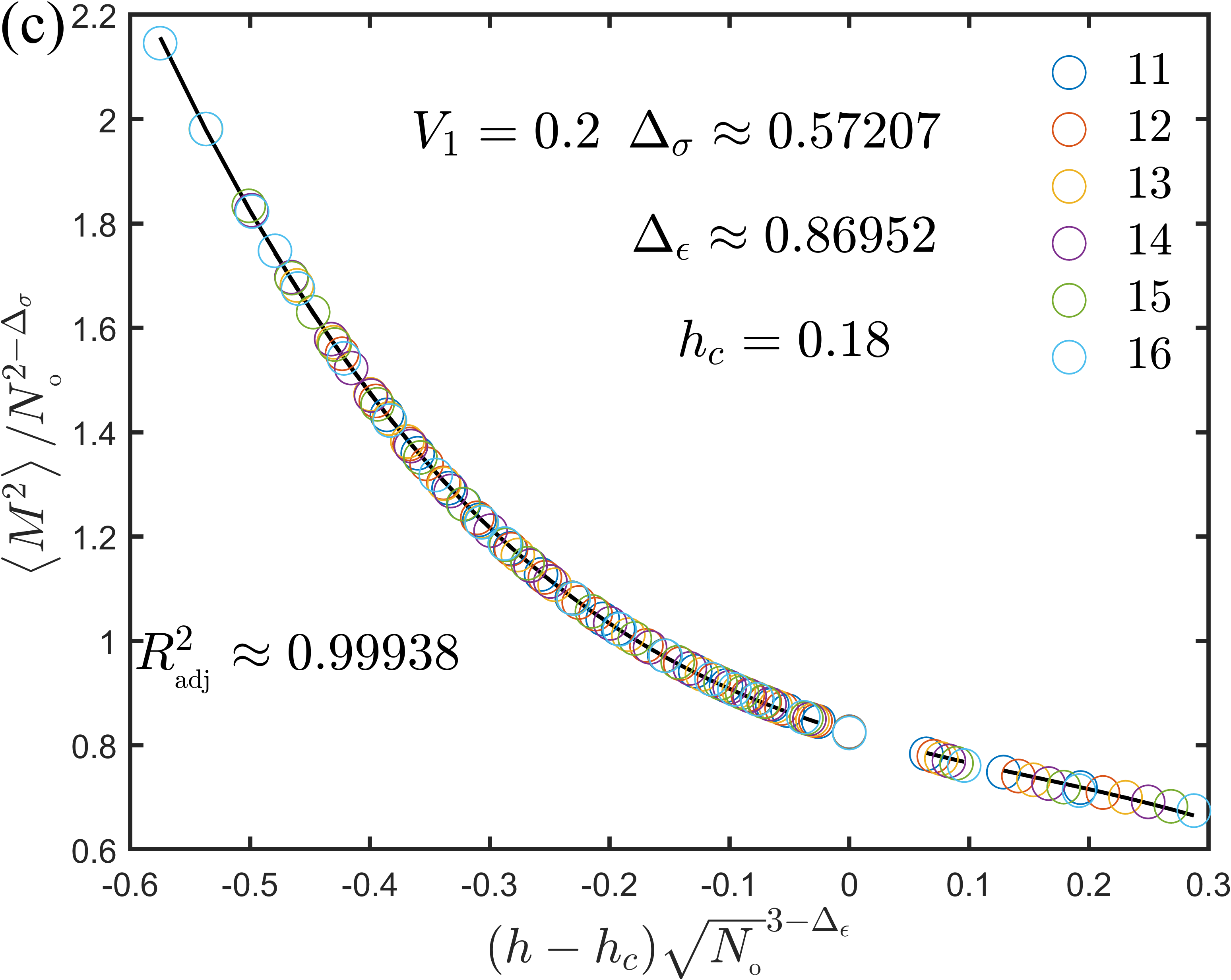}
\includegraphics[width=0.49\linewidth]{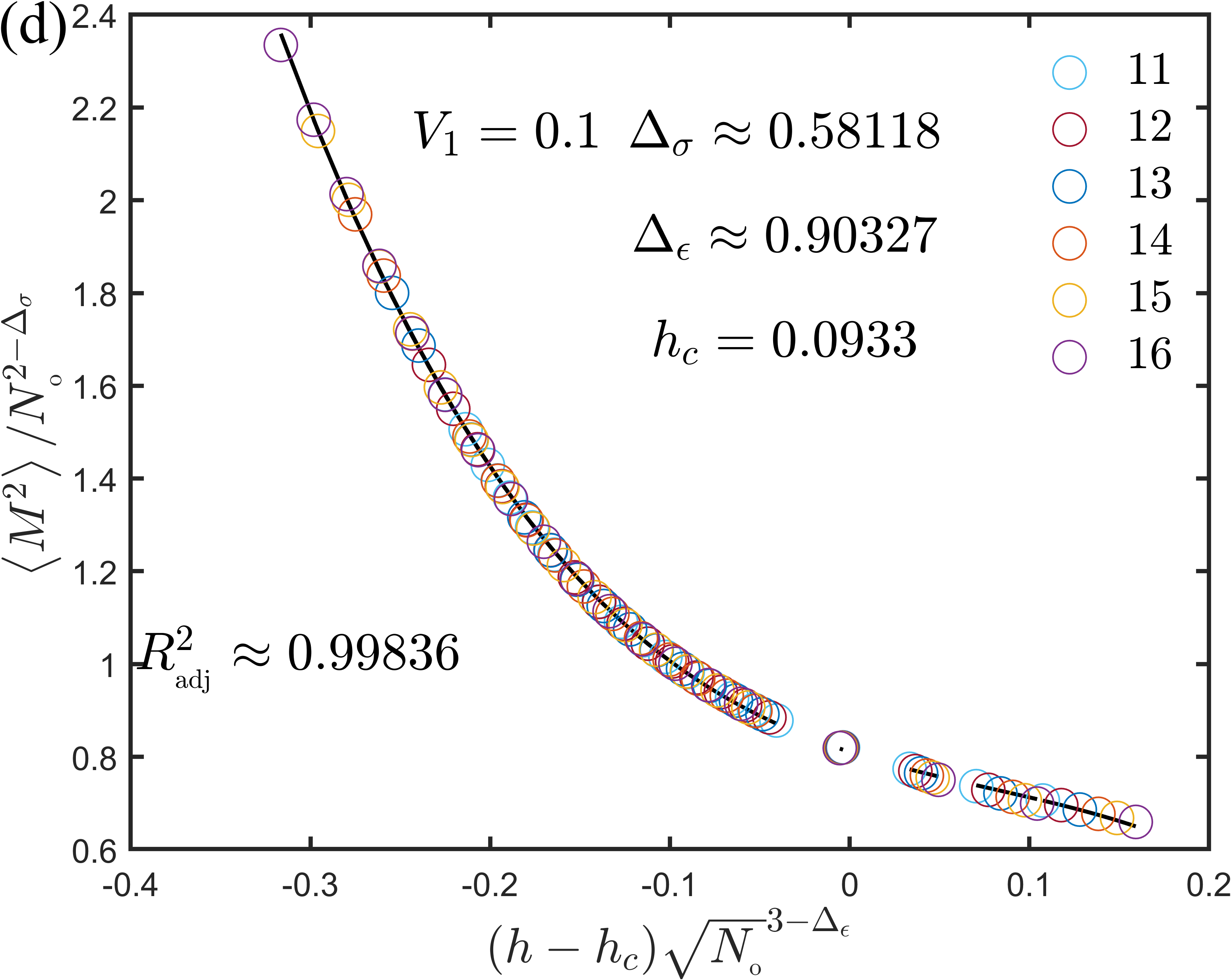}
	\caption{ 
    (a) Binder ratio $U_4$ versus $h$ for $V_1=0.2$. (b) The extrapolation for the crossing point of two successive system sizes$(N_{\mathrm{o}},N_{\mathrm{o}}+1)$.  We fit the data through $h_c(N_{\mathrm{o}})=aN_{\mathrm{o}}^{-x}+h_c$, which gives $h_c\approx 0.18$ at thermodynamic limit.  The numerical results for $N_{\mathrm{o}}=4\sim 8$ are computed by the ED, and results for $N_{\mathrm{o}}=9\sim 16$ are computed using DMRG. (c)(d) Finite-size scaling of the recaled order parameters. The data collapse is achieved by assuming $\left\langle M^2\right\rangle/N_{\mathrm{o}}^{2-\Delta_\sigma}$ depends on the function form  $f_0 + f_1 x + f_2 x^2 + f_3 x^3$, where $x = (h - h_c)\sqrt{N_\mathrm{o}}^{3-\Delta_\epsilon}$, with the phase transition point \( h_c \) fixed, the parameters \( f_0 \), \( f_1 \), \( f_2 \), \( f_3 \), \( \Delta_\sigma \), and \( \Delta_\epsilon \) are fitting parameters.The adjusted coefficient of determination \(R^2_{adj}\rightarrow 1\) indicates that the fitting has successfully captured data features while avoiding overfitting. 
    } 
	\label{fig:U4}
\end{figure}


\section{Order parameter analysis}
We firstly discuss the magnetic order parameter and use it
to determine the phase diagram. We consider the following order parameter to characterize the spontaneous breaking of the $S_3$ symmetry:
\begin{equation}
    M=\sum_{m=-s}^{s}\boldsymbol{c}_m^\dagger S_z \boldsymbol{c}_m.
\end{equation}
If the phase transition belongs to the continuous type, at the phase transition point, this $S_3$ order parameter should be scaled as 
\begin{align}\label{eq:orderparameter}
\left\langle M^2\right\rangle \sim R^{4-2 \Delta_\sigma}\sim N_{\mathrm{o}}^{2-\Delta_\sigma} =N_{\mathrm{o}}^{2-(1+\eta)/2},
\end{align}
where the typical length on fuzzy two-sphere is proportional to the square root of the number of Landau orbitals by $R \sim \sqrt{N_{\mathrm{o}}}$ and $\Delta_\sigma$ is the scaling dimension of magnetic order parameter. The critical exponent is $\eta=2\Delta_\sigma-1$ for $(2+1)$-D space-time. By analyzing this scaling behavior, we can gain insights into the critical properties of the phase transition and the underlying symmetry-breaking mechanisms. 
In particular, we examine the RG invariant Binder ratio $ U_4=\left <M^4\right>/\left<M^2\right>^2$\cite{1981ZPhyB..43..119B}, focusing on its finite-size scaling around the phase transition to confirm the estimation of $h_c$.

Fig.~\ref{fig:U4}(a) illustrates the Binder cumulant $U_4$ as a function of $h$ for various system sizes $N_{\mathrm{o}}$. The plot indicates that at small values of $h$, the model resides in the ordered phase, characterized by large values of spontaneous magnetization. Conversely, at large values of $h$, the model transitions into the disordered phase, where the magnetization vanishes. There is a crossing region $h\approx h_c$, where different system sizes cross with each other, signaling the phase transition point. Fig.~\ref{fig:U4}(c) depict the plot of $\left\langle M^2\right\rangle/N_{\mathrm{o}}^{2-\Delta_\sigma}$ versus $h$ of different system sizes $N_{\mathrm{o}}$ with $V_1=0.2$. We find that all data points nicely collapse onto a single curve near $h_c\approx0.18$.
Based on the same analysis, we obtain a global phase diagram, Fig. \ref{fig:phasediagram}, for various interaction parameters $V_1$. 
Next, we will analyze the location of transition point $h_c$ from different angles.

\begin{figure}[t]
\includegraphics[width=0.49\linewidth]{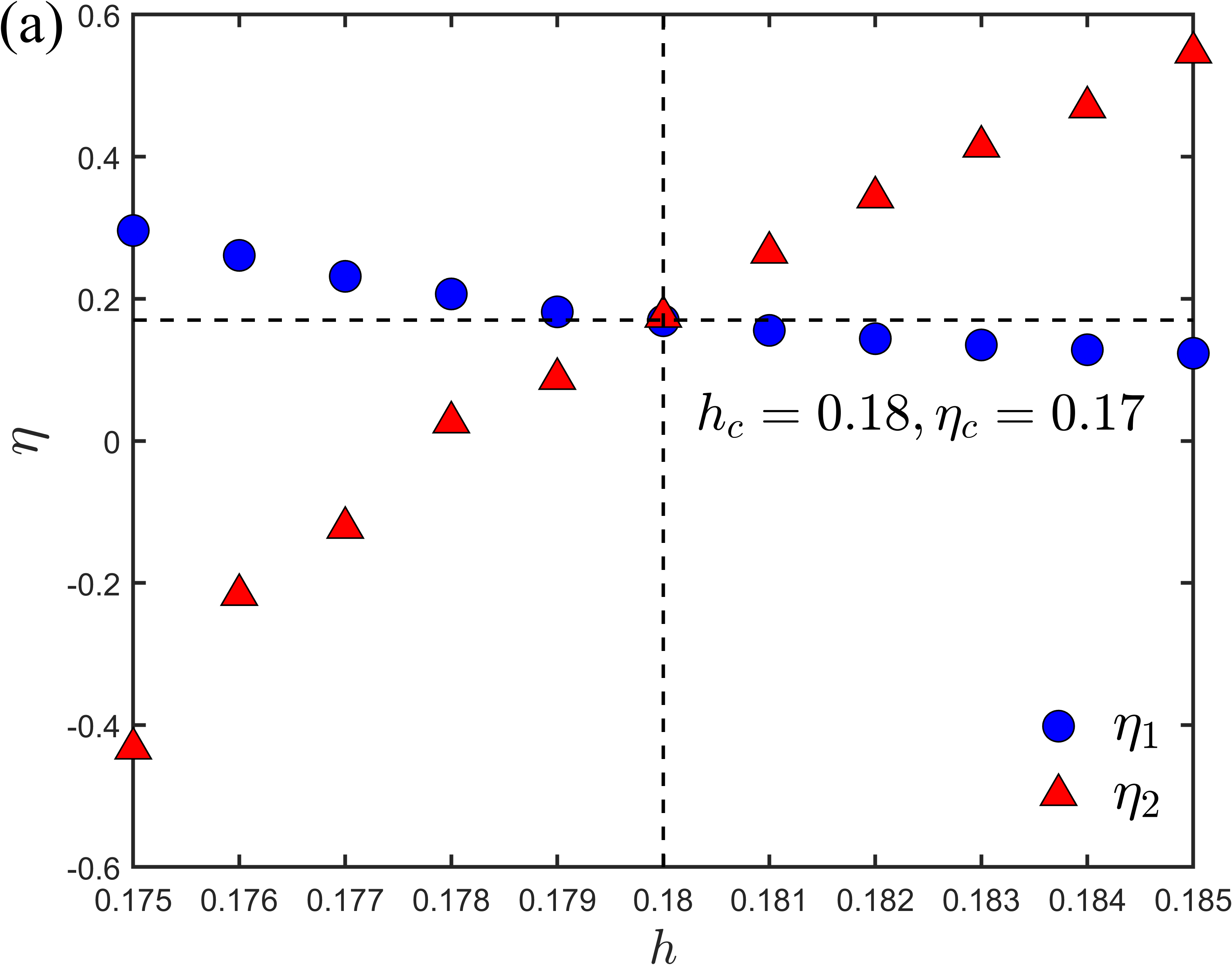}
\includegraphics[width=0.49\linewidth]{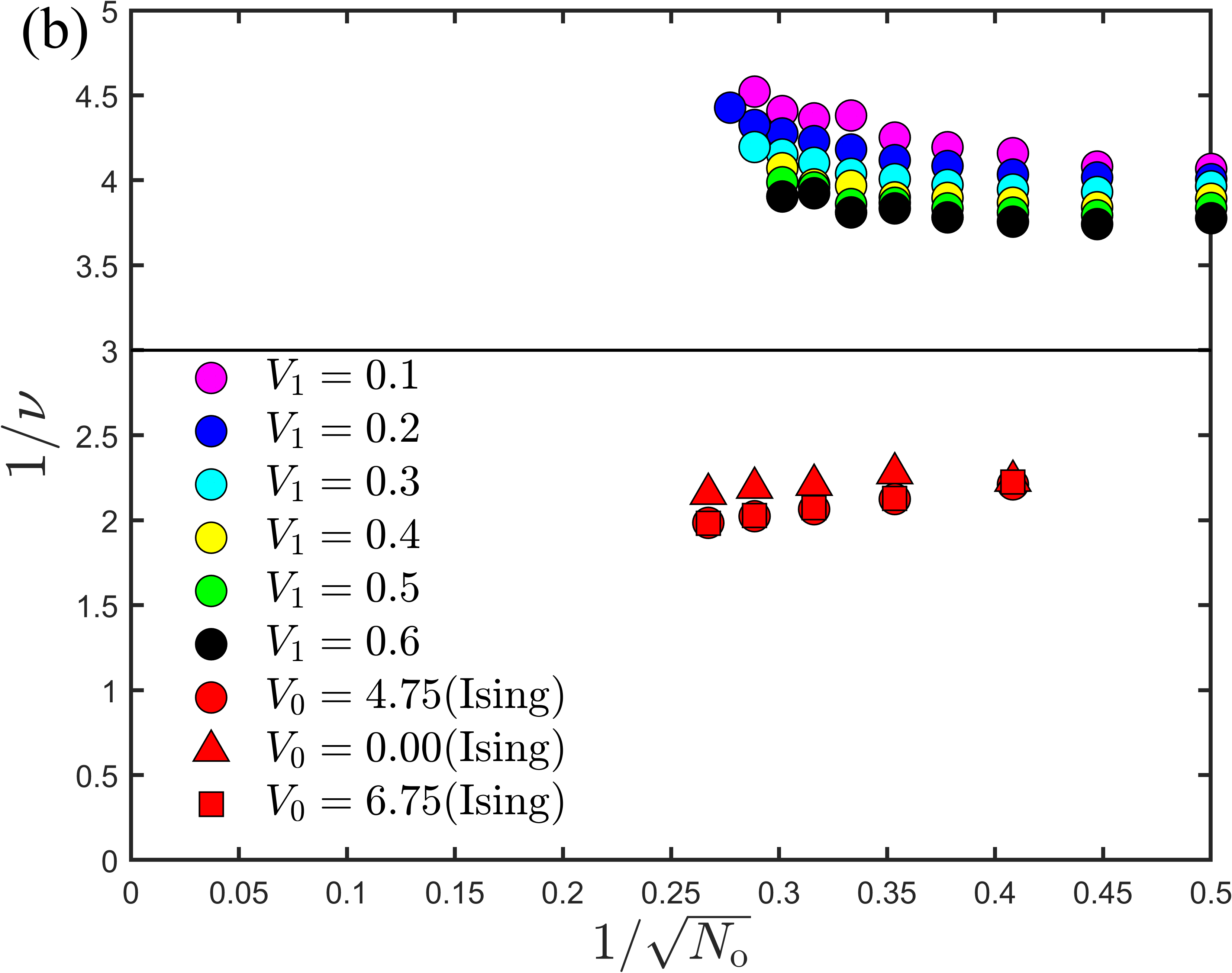}
	\caption{
 (a) The extrapolated values $\eta$ near the estimated transition point obtained from fitting the order parameter using Eq. \ref{eq:eta_1} (blue dots) and Eq. \ref{eq:eta_2} (red dots). The consistency check indicates the transition occurs around $h_c\approx 0.18$ for $V_1=0.2$.
 (b) The flow of inverse critical exponent $1/\nu(N_o)$ with inverse systems size $1/\sqrt{N_{\mathrm{o}}}$, for $V_1/V_0=0.1,0.2,0.3$. As a comparison, we also show the case for the 3D Ising model.
  }
	\label{fig:eta_extrapolate}
\end{figure}

\subsection{Determination of the transition point}
Since this $2+1$-D Potts phase transition has been less studied before, all critical exponents remain unknown, making it essential to determine the transition point accurately. First, an estimate of the
value of critical point $h_c$ can be determined by the crossing-point analysis according to the scaling form $h_c\left(N_{\mathrm{o}}\right)=a N_{\mathrm{o}}^{-x}+b$, as shown in Fig.~\ref{fig:U4}(b). These data points are given by the crossing points of the binder ratio for different successive size pairs $(N_\mathrm{o},N_\mathrm{o}+1)$ crosses around the point $h_c\approx0.18$.

Next, to further confirm the validity of the transition point $h_c\approx0.18$, we propose the following self-consistent check based on the critical exponent $\eta$. 
According to Eq. \ref{eq:orderparameter}, we can consider the finite-size scaling of the reduced order parameter ($\left<m^2\right>=\left<M^2\right>/N_{\mathrm{o}}^2$) 
\begin{equation}
\label{eq:eta_1}
    \left<m^2(N_{\mathrm{o}})\right>=m^2(\infty)+aN_{\mathrm{o}}^{-(1+\eta)/2}
\end{equation}
to describe the size dependence of the order parameter near the transition point, from which an estimate of the critical exponent $\eta$ can be obtained. On the other hand, 
an alternative approach to extract $\eta(N)$ on finite-sizes employs the following formula \cite{PhysRevLett.121.117202}
\begin{equation}
    \eta(N_{\mathrm{o}})=-\frac{1}{\ln r}\ln\left[\frac{m^2(h_c,N_{\mathrm{o}}+x)}{m^2(h_c,N_{\mathrm{o}})}\right]-1,
\end{equation}
where $r=\sqrt{\frac{N_{\mathrm{o}}+x}{N_{\mathrm{o}}}}$. By fitting the finite size values through:
\begin{equation}
\label{eq:eta_2}
    \eta(N_{\mathrm{o}})=\eta(\infty)+N_{\mathrm{o}}^{-\omega/2},
\end{equation}
we can extract the value of $\eta$ in the thermodynamic limit. The exponents $\eta$ determined by both methods (Eq. \ref{eq:eta_1} and Eq. \ref{eq:eta_2}) should be consistent near the transition point. The Fig. \ref{fig:eta_extrapolate}(a) shows the values of the critical exponent $\eta$ extracted using two different methods,  for \(V_1 = 0.2\). It can be seen that the two are consistent only around \(h_c \approx 0.18\), while for values of \(h\) deviating from \(0.18\), there is a significant discrepancy between the $\eta$ values obtained from the two methods. Based on this, we identify \(h_c = 0.18\) as the ferromagnetic-to-paramagnetic transition point for \(V_1 = 0.2\).
Moreover, at $h_c=0.18$ it is found $\eta\approx 0.17$, which gives the scaling dimension $\Delta_\sigma = \frac{\eta+1}{2} \approx 0.585$, this is roughly consistent with the conformal data estimated by the operator spectrum (see below).

\subsection{Evidence of first-order transition}
Moreover, we can extract the critical exponent $\nu$ regarding the phase transition based on the binder ratio. 
Here, the finite-size value \(\nu(N_o)\) is extracted from the slope of the Binder ratio at the intersection points of two consecutive different sizes, based on the following equation \cite{doi:10.1126/science.aad5007,PhysRevB.31.3069}: 
\begin{equation}
    \frac{1}{\nu (N_\mathrm{o})}=\frac{1}{\ln{r}}\left[\ln{\frac{\frac{dU_4}{dh}(h,N_{\mathrm{o}}+x)}{\frac{dU_4}{dh}(h,N_{\mathrm{o}})}}\right]_{h=h_{\mathrm{cross}}({N_{\mathrm{o}}})}
\end{equation}
where \(h_{\mathrm{cross}}(N_{\mathrm{o}})\) is the intersection of the Binder ratio for two different sizes \((N_{\mathrm{o}}, N_{\mathrm{o}}+x)\). The derivatives are obtained by interpolating the dense data near the crossing point of two sizes. Thus, the inverse correlation length critical exponent in the thermodynamic limit can be fitted using $\frac{1}{\nu (N_{\mathrm{o}})}=\frac{1}{\nu (\infty)}-cN_{\mathrm{o}}^{-\frac{\omega}{2}}$. 
Fig. \ref{fig:eta_extrapolate}(b) shows that the exponent $1/\nu$ flows with the system size $N_o$ towards an unphysical value ($>3$), which implies the transition is not standard second-order type \cite{Iino2019}. 
In contrast, for the case of 3D Ising universality class \cite{ZHHHH2022}, the size-dependent \(1/\nu\) obtained in the same manner changes more smoothly, with the extrapolated value still being less than the space-time dimension-3. 
Therefore, the value and drifting behavior of $\nu$ serve as evidence of first-order transition for 3D Potts transition.

To sum up, we present several physical observables regarding the magnetic order parameter across the phase transition from Potts ferromagnetic phase to the paramagnetic phase. We find that simply applying the data analysis schemes, which are widely applied to the second-order phase transitions,  
to study the 3D Potts transition inevitably leads to some anomalous behaviors, i.e., critical exponent $1/\nu$ violates the physics bound. 
This "anomalous" behavior implies that the 3D Potts transition is of the first-order type instead of the continuous type.  

Before ending this section, we stress that the data analysis in this section is based on the assumption that the phase transition is the second-order type. These results should be treated with caution. For example, the data collapse in Fig. \ref{fig:U4}(c-d) looks perfect. However, it does not imply the second-order phase transition directly. One plausible reason could be the first-order transition with a system size much smaller than the correlation length ($L < \xi$). Or, the observation in Fig. \ref{fig:U4}(c-d) does not contradict the first-order transition (more evidence see below).

\begin{widetext}
\begin{figure*}[ht!b]
\centering
\includegraphics[width=1.0\textwidth]{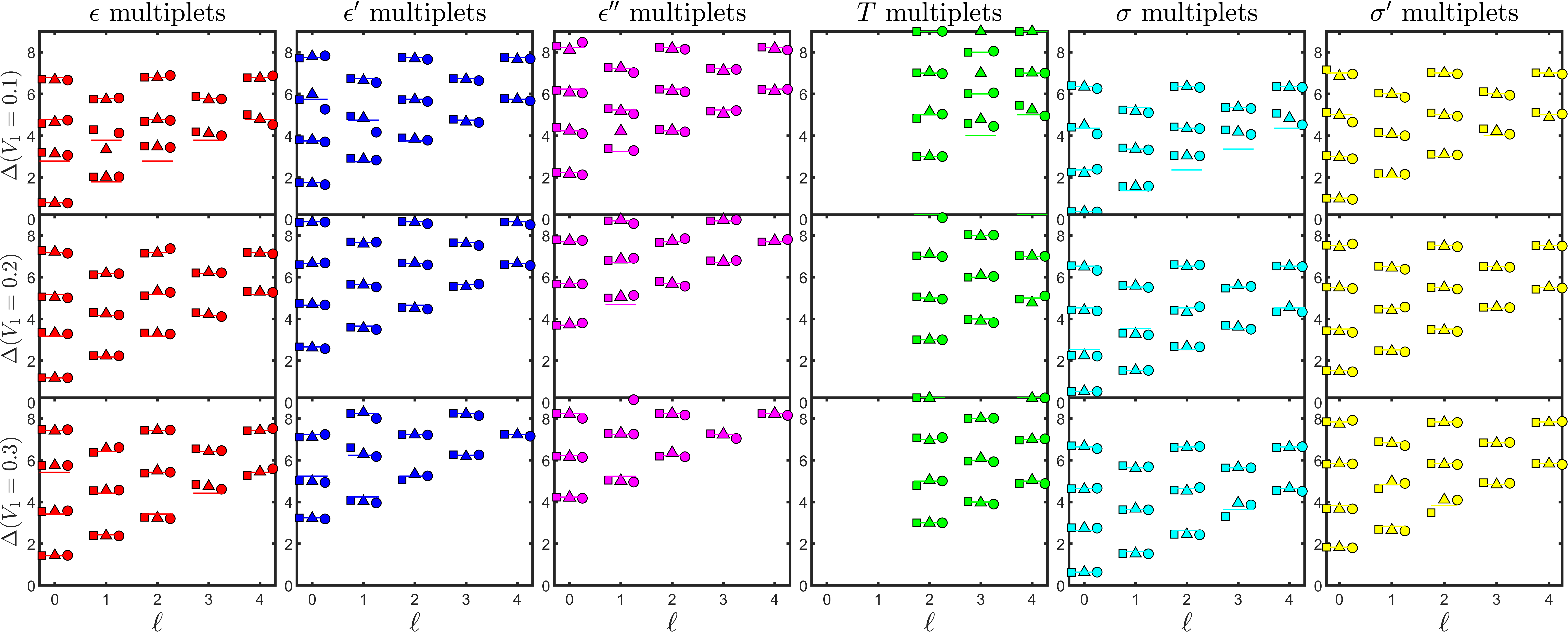}
\caption{ Conformal multiplet for primary operators: scaling dimension $\Delta$ versus Lorentz spin $\ell$. Different colors label different conformal towers. The spectrum is calibrated by setting the scaling dimension of energy-momentum tensor $\Delta_T=3$.  The dots are results from numerical calculation, $V_1=0.1,h_c=0.0933,V_1=0.2,h_c=0.18,V_1=0.3,h_c=0.263$. Different symbols denote different total system sizes (circle: $N_{\mathrm{o}}=6$, triangle: $N_{\mathrm{o}}=7$, square: $N_{\mathrm{o}}=8$).
}\label{fig:con_fam}
\end{figure*}
\end{widetext}

\section{Operator spectrum}
On the fuzzy sphere, the eigenenergy gaps of the quantized critical Hamiltonian have a one-to-one correspondence with the scaling dimensions of the underlying CFT operators, dubbed as the state-operator correspondence \cite{Cardy1984,Cardy1985a}.  
Following the discussion in Ref. \cite{ZHHHH2022}, we choose the energy-momentum tensor $T_{\mu\nu}$ as the calibrator to rescale the energy spectrum. This is because for any $3D$ CFT, $T_{\mu\nu}$ is a conserved operator, which should be a primary with Lorentz spin $\ell=2$ and scaling dimension $\Delta_T=3$.  
In particular, a typical feature of the conformal symmetry is the integer-spaced levels, i.e. for a given CFT primary $\mathcal{O}$ with quantum numbers $(\ell,\mathrm{rep.},...)$ and scaling dimensions $\Delta_{\mathcal{O}}$, its descendants should have scaling dimensions $\Delta_{\mathcal{O}}+n$ ($n$ is positive integer) with definite quantum numbers determined by the symmetry operations (see e.g. Ref.~\cite{ZHHHH2022} for a detailed discussion). 

In the rescaled operator spectrum Fig. \ref{fig:con_fam}, we observe that the low-lying levels indeed exhibit an approximate integer-spaced pattern as required by the conformal symmetry. Fig.~\ref{fig:con_fam} shows the numerically identified conformal multiplet (i.e. primary and its descendants) of the lowest two $\mathrm{S}_3$ vector operators $\sigma,\sigma'$, the lowest three $\mathrm{S}_3$ singlet operators $\epsilon,\epsilon',\epsilon''$ and the energy-momentum tensor $T_{\mu\nu}$ by matching the quantum numbers. We can find their descendants up to $\ell\le 4$ and $\Delta\le 9$. The measured scaling dimensions (symbols) and the corresponding anticipated values (solid lines) largely agree with each other. Many intervals are not precisely integer-spaced in Fig. \ref{fig:con_fam}. This may be attributed to the lack of exact conformal symmetry due to the pseudo-criticality and finite-size effect (see the discussion below). 
These results convincingly demonstrate the emergent conformal symmetry despite not being exact. 
Furthermore, we also show three critical parameters along the transition line for $V_1=0.1, 0.2 ,0.3$ with different system sizes $N_\mathrm{o}$. It is also worth noting that this emergent symmetry exists along the phase transition line, which hints at the exotic physics behind it. 
Additionally, we can define a cost function to quantify the deviation of data away from the conformal theory.
Our extensive search shows that the conformal tower structure is best around $V_1=0.2$. Therefore, we will mainly focus on this parameter to present our further analysis.

\begin{table}[b]
\caption{\label{tab:ope_dim_primary} Operator scaling dimensions for primary fields identified through the state-operator correspondence. $\ell$ represents the Lorentz spin quantum number. $\mathrm{rep}$ labels the $S_3$ conformal group representation. These data are calculated at the transition point $h_c=0.18$ with $V_1=0.2$ by setting $\Delta_{T_{\mu\nu}}=3$. 
}

\begin{ruledtabular}
\begin{tabular}{ccccccccccc}
Op.  & $\ell$ &$\mathrm{rep}$ & $14$ & $13$ & $12$ & $11$ & $10$ & $9$ & $8$ \\ \hline

$\epsilon$ &  $0$ &$\boldsymbol{0}^+$ & $1.1622$ & $1.1647$ &$1.1671$  &$1.1693$ &$1.1714$  &$1.1732$ & $1.1746
$     \\

$\epsilon'$ & $0$ &$\boldsymbol{0}^+$ & $2.7468$ & $2.7406$ &$2.7321$  &$2.7205$ & $2.7050
$ & $2.6845
$ & $2.6573
$   \\

$\epsilon''$  & $0$ &$\boldsymbol{0}^+$ & $-$ & $-$ &$3.6825$  &$3.6728
$ & $3.6698 
$ & $3.6770
$ & $3.6983
$ \\ \hline

$\sigma$  & $0$ &$\boldsymbol{1}$ & $0.5439$ & $0.5421$ &$0.5401$  &$0.5378 
$ & $0.5353$ & $0.5325
$ & $0.5293
$  \\

$\sigma'$  & $0$ &$\boldsymbol{1}$ & $1.5957$ & $1.5869$ &$1.5767$  &$1.5648$ & $1.5509$ & $1.5343
$ & $1.5146
$  \\ 

\end{tabular}
\end{ruledtabular}
\end{table}

\begin{figure}[t]
	\centering
    \includegraphics[width=0.5\textwidth]{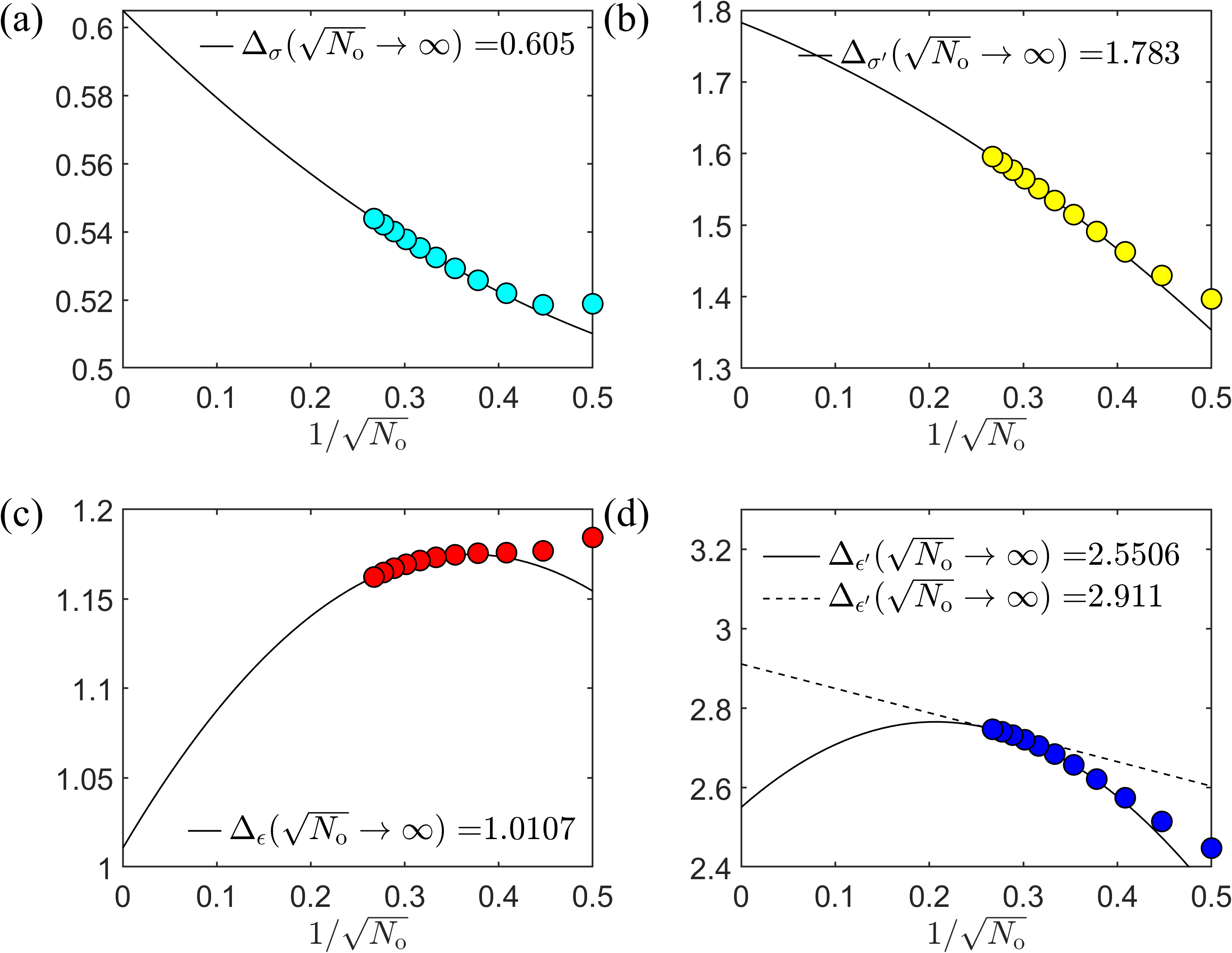}
	\caption{	
        Finite-size extrapolations of the scaling dimensions of (a) $\Delta_\sigma$  (b) $\Delta_{\sigma^\prime}$ (c) $\Delta_\epsilon$  and (d) $\Delta_{\epsilon'}$. These operators are calculated at the transition point $h_c=0.18$ with $V_1=0.2$. We use the form of $\Delta(N_{\rm{o}})=a/N_{\rm{o}}+b/\sqrt{N_{\rm{o}}}+\Delta(\infty)$ to obtain the scaling dimension of the operator in the thermodynamic limit (note the typical length $R \sim \sqrt{N_o}$).
        In subfig (d), we use both linear (dashed, only largest two sizes used) and polynomial (solid) scaling form (using data with size $N_{\mathrm{o}} > 8$). 
	}
	\label{fig:scaling}
\end{figure}

Since our calculations are only available to limited number of electrons, it is necessary to analyze the finite-size effect in detail. 
First, in Fig. \ref{fig:scaling} we plot the size dependence of the scaling dimensions of the lowest four primaries $\sigma,\sigma',\epsilon,\epsilon'$, and the related data are shown in Tab. \ref{tab:ope_dim_primary}. All scaling dimensions drift with the system size $R\sim \sqrt{N_{\mathrm{o}}}$.
Most of them have non-linear dependence on the system size, and it is difficult to analyze the extrapolated values in the thermodynamic limit $R\rightarrow \infty$. Although extracting reliable extrapolated values is difficult, we can obtain some qualitative conclusions at this step.  
For example, in Fig. \ref{fig:scaling}(a), the scaled magnetic scaling dimension $\Delta_\sigma \rightarrow 0.60$. This value is close to the estimation ($\Delta_\sigma\approx 0.585$) using the order parameter in Fig. \ref{fig:eta_extrapolate}.  

Moreover, since the available system sizes are limited, we could not precisely determine the scaling dimension $\Delta_{\epsilon'}$ for the second lowest $S_3$ singlet primary $\epsilon'$. 
Fig. \ref{fig:scaling}(d) shows that $\Delta_{\epsilon'}$ drifts from relevant to dangerously relevant with the increase of $N_{\mathrm{o}}$. We get $\Delta_{\epsilon'} (N_{\mathrm{o}} \to \infty)\approx 2.911$ using the linear fit, and $\Delta_{\epsilon'}(N_{\mathrm{o}} \to \infty)\approx 2.55$ using the polynomial scaling. 
However, we need to emphasize the necessity for caution when using these scaled values. On the one hand, if the operator $\epsilon'$, with regard to the nearby conformal fixed point, is relevant, the long-wavelength limit of most phase transition points in the phase diagram will flow away from this fixed point and it is unreasonable to use the conformal perturbation theory directly to perform the extrapolation. Within this case, one should include further interactions into the Hamiltonian and inspect an extended Potts model to determine the exact continuous phase transition point.  On the other hand, even if the operator $\epsilon'$ is dangerously irrelevant, the third scalar primary $\epsilon''$ has the scaling dimension around $\Delta_{\epsilon''}\approx 3.68$ from our numerical calculation, see Tab. \ref{tab:ope_dim_primary}. So we expect the lowest finite-size correction scales like $\Delta(N_{\mathrm{o}}) \sim \Delta(\infty)+\frac{a} {R^{(\Delta_{\epsilon'}-3)}}+\frac{b}{R^{(\Delta_{\epsilon''}-3)}}$, i.e. both power indexes are fractional values less than 1 instead of an integer value. Nevertheless, if we fit our raw data with this scaling formula, the extrapolated results change severely and lead to some unphysical values. This indicates that the renormalization group flows are much slower within our model, potentially due to the nearby multicritical/complex fixed point (see discussion below). In this context, neither linear nor polynomial function is the right choice.

\begin{figure}
    \centering
    \includegraphics[width=0.49\linewidth]{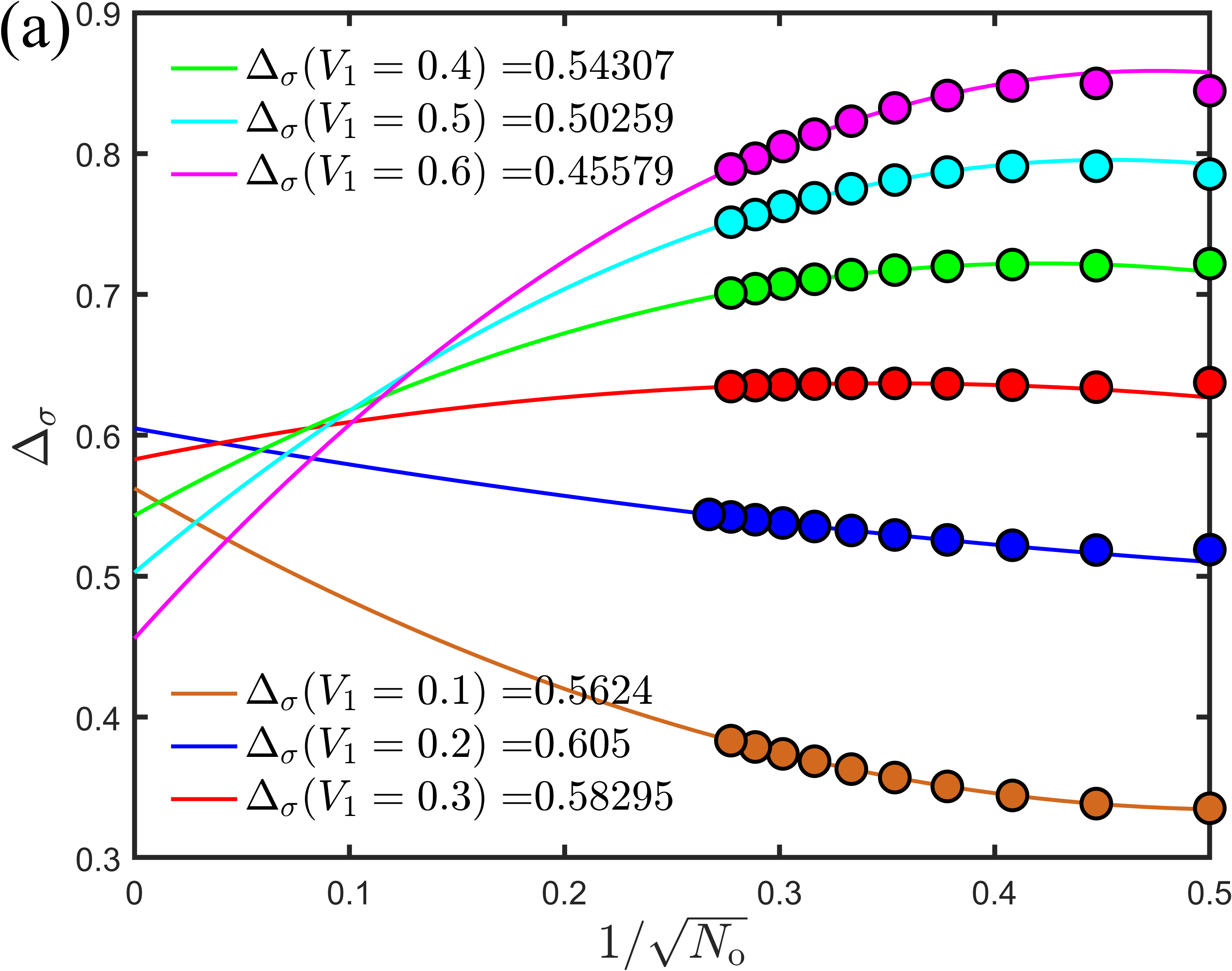}
    \includegraphics[width=0.49\linewidth]{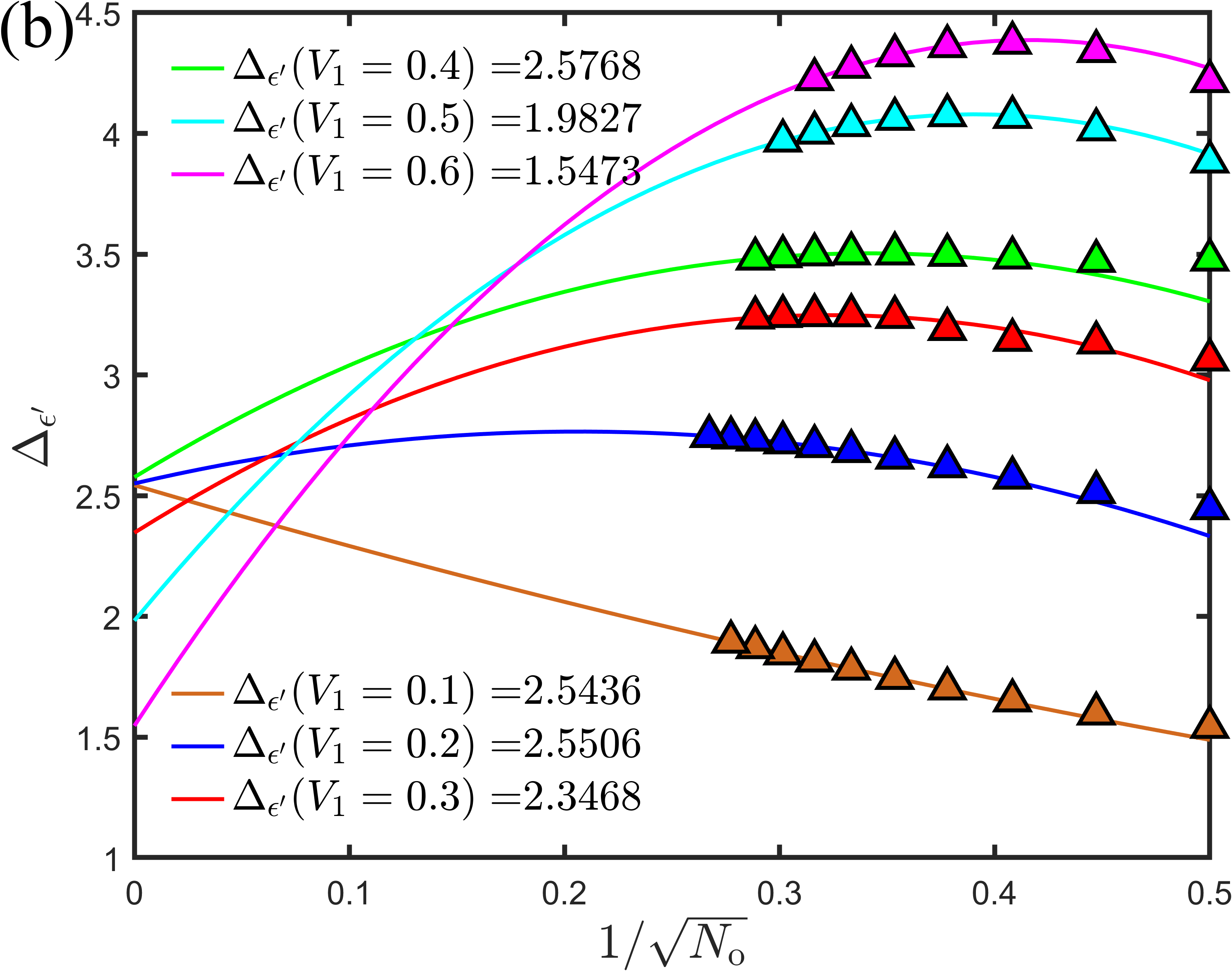}
    \includegraphics[width=0.49\linewidth]{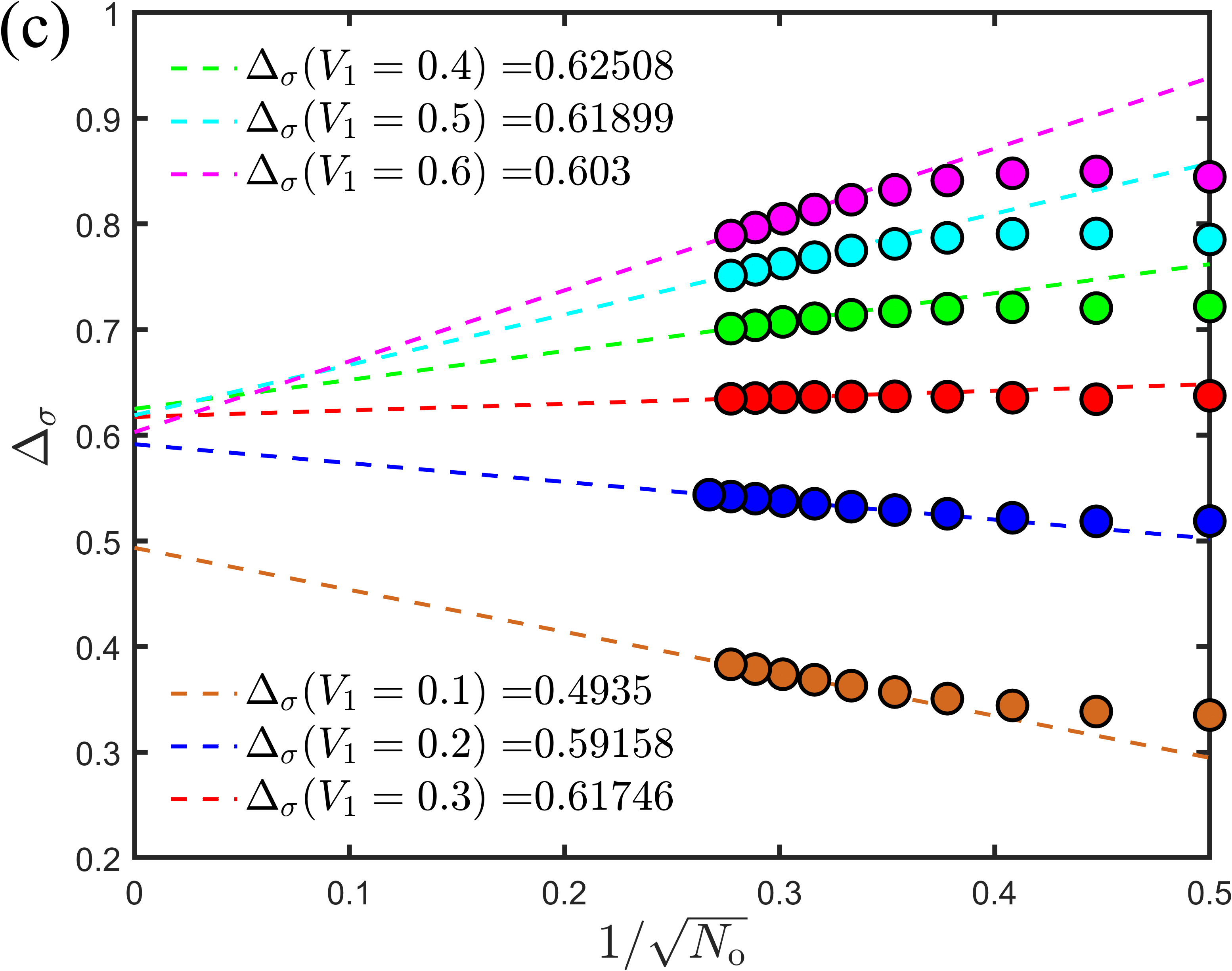}
    \includegraphics[width=0.49\linewidth]{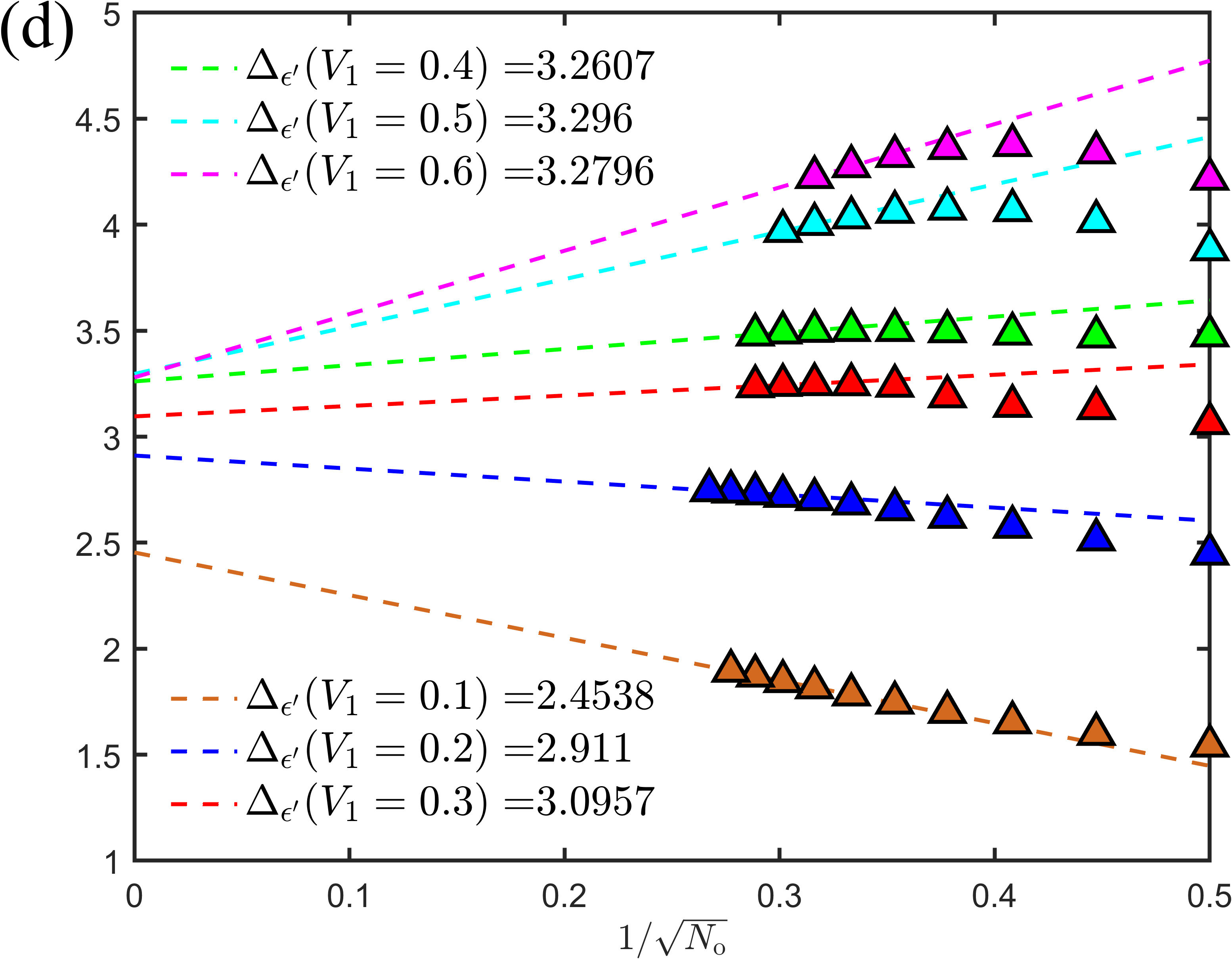}
    \caption{ Comparison of finite-size extrapolations of (a,c) $\Delta_{\sigma}$ and (b,d) $\Delta_{\epsilon'}$ for various transition points $V_1=0.2-0.6$ (see Fig. \ref{fig:phasediagram}). 
    In (a) and (b), the polynomial function (solid line) is used for finite-size scaling. 
    In (c) and (d), the linear function (dashed line) is used.
    }
    \label{fig:comp_finitesizescaling}
\end{figure}

Tab. \ref{tab:diff_V1_primary} presents more data for $V_1=0.1-0.6$. Fig. \ref{fig:comp_finitesizescaling} depicts the scaling dimension of $\Delta_{\sigma}, \Delta_{\epsilon'}$ for various transition points and compares different scaling forms. We find that the convergent trend is roughly the same across various transition points. For the linear fits in Fig. \ref{fig:comp_finitesizescaling} 
 (c-d), $\Delta_{\sigma}$ approximately approaches $\sim 0.6$, and $\Delta_{\epsilon'}$ gives a value close to $\sim 3$.  
Compared to the linear fits, the range of scaled values fitted via the polynomial functions is larger. Another feature is, in the polynomial fit, $\Delta_{\epsilon'}$ is clearly below 3 (Fig. \ref{fig:comp_finitesizescaling}(b)). 
Since the extrapolated value of $\Delta_{\epsilon'}$ is sensitive to the fitting process and Hamiltonian parameters, the precise value of $\Delta_{\epsilon'}$ in the thermodynamic limit is still uncertain. 
Nevertheless, since the data clearly show non-linear behaviors, we are inclined to accept the result from the polynomial scaling. 

Here, we would like to make some remarks. 
We suspect $\epsilon'$ is highly relevant to the appearance of the first-order transition.
Let us recall the 2D critical Potts model, where $\epsilon'$ is irrelevant for $Q\le Q_c(2)=4$ while it becomes relevant for $Q>4$ \cite{Cardy1980},  $\epsilon'$ drives ``walking" RG flow and induces the first-order transition in the original Potts model \cite{Gorbenko2018a,Gorbenko2018b}. In the 3D Potts model, from our perspective, identifying the relevance of  $\epsilon'$ is essential for the nature of transition, which could indicate how the merge-and-annihilate picture works.

\textit{}
\begin{table}[]
\caption{ Scaling dimensions of several primary fields extracted at the phase transition points for different $V_1/V_0$. 
}\label{tab:diff_V1_primary}
\begin{tabular}{c|c|ccccccc}
\hline\hline
\multirow{2}{*}{Op.}       & \multirow{2}{*}{$V_1/V_0$} & \multicolumn{7}{c}{$N_{\text{o}}$} \\ \cline{3-9} 
                                &       & $13$      & $12$       & $11$        & $10$   & $9$   & $8$  & $7$    \\ \hline
\multirow{6}{*}{$\epsilon$}     
& 0.1 &0.8140  &0.8092   &0.8038    &0.7978    &0.7911    & 0.7837    &0.7756   \\
                                & 0.2   &1.1647  &1.1671   &1.1693     & 1.1714    & 1.1732   &1.1745    &1.1754     \\
                                & 0.3    &1.3740   &1.3832      &1.3928      & 1.4028     &1.4131     & 1.4235   & 1.4336     \\ 
                                & 0.4    &1.5234  &1.5385    &1.5548   &1.5720      &1.5901  & 1.6089    &1.6274   \\
                                & 0.5     &1.6356  &1.6562   &1.6785   & 1.7026     & 1.7282   &1.7550    &1.7820     \\
                                & 0.6      &1.7203    &1.7460   &1.7740        & 1.8043     &1.8370     & 1.8716   & 1.9067     \\ \hline
\multirow{6}{*}{$\epsilon'$} & 0.1       &1.8954    &1.8726     &1.8467        & 1.8173     & 1.7836    & 1.7452   & 1.7015     \\
                                & 0.2       &2.7406    &2.7321     &2.7205        & 2.7050     & 2.6845    & 2.6573   & 2.6215     \\
                                & 0.3    &   &3.2374  &3.2437    & 3.2473     & 3.2468   & 3.2398   & 3.1899    \\ 
     & 0.4     &      &3.4812      &3.4911       & 3.4987     & 3.5029    & 3.5027   & 3.4970     \\
                                & 0.5       &   &    &3.9701    & 4.0030    & 4.0342    & 4.0603   & 4.0753     \\
                                & 0.6      &     &   &    & 4.2243     & 4.2754   & 4.3234   & 4.3611    \\ \hline
\multirow{6}{*}{$\sigma$}       & 0.1       &0.3833      &0.3788     &0.3739      & 0.3687     & 0.3631    & 0.3571   & 0.3508     \\
                                & 0.2          &0.5421   &0.5401      &0.5378     & 0.5353    & 0.5325   & 0.5293   & 0.5258    \\
                                & 0.3          &0.6346    &0.6353      &0.6359    & 0.6364     & 0.6368   & 0.6369   & 0.6366    \\ 
                                 & 0.4         &0.7010    &0.7041       &0.7074    & 0.7108     & 0.7141    & 0.7173   & 0.7199     \\
                                & 0.5          &0.7513     &0.7567     &0.7625    & 0.7686    & 0.7750   & 0.7812   & 0.7869    \\
                                & 0.6          &0.7891      &0.7967    &0.8050    & 0.8138     &  0.8230  & 0.8324   & 0.8412    \\ \hline\hline
\end{tabular}
\end{table}

\section{ Correlation functions and OPE coefficients}
\label{sec:ope}
Next, we will continue to investigate the correlation functions \cite{Four_Han2023} and the operator product expansion (OPE) coefficients \cite{OPE_hu2023} in the context of the corresponding conformal field theory.

We chose the local lattice operator \( \hat{n}_z(\Omega) \) to approximate the primary field \(\sigma\), and \( \hat{n}_x(\Omega) \) or the following operator \( \hat{O}_\epsilon(\Omega_a) \) to approximate \(\epsilon\), 
 \begin{equation}
 \begin{aligned}
     \hat{O}_\epsilon(\Omega_a)=&\int{d\Omega_bU\left( \Omega_{ab} \right)\left[ n_0\left( \Omega_a \right) n_0\left( \Omega_b \right) -n_z\left( \Omega_a \right) n_{z}^{\dag}\left( \Omega_b \right) \right]}\\
&+h\left[ n_x\left( \Omega_a \right) +n_{x}^{\dag}\left( \Omega_a \right) \right]
 \end{aligned}
 \end{equation}
The only difference between this operator and the local Hamiltonian operator is that the coefficient of the transverse field term is inverted. For computational convenience, we often utilize the spherical modes of these local operators defined in the orbital space.
\begin{equation}
    \hat{O}_{l,m}=\int d\Omega \Bar{Y}_{l,m}(\Omega)\hat{O}(\Omega)
\end{equation}

\begin{figure}
    \centering
    \includegraphics[width=0.9\linewidth]{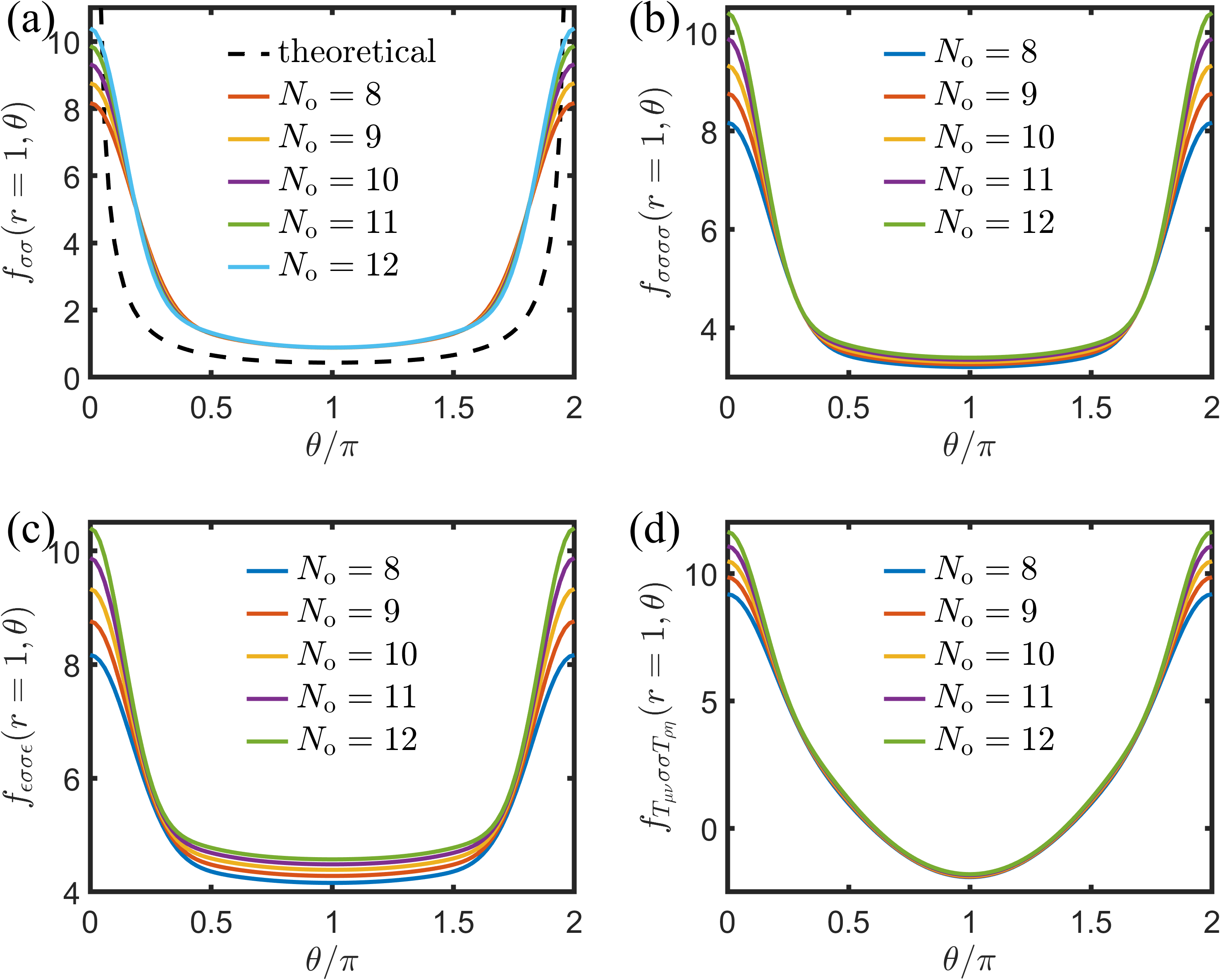}
    \caption{ The angle
 dependence of (a)two-point correlator $f_{\sigma\sigma}(r=1,\theta)$, The dashed line is plotted according to the theoretical formula \( (2\sin{\frac{\theta}{2}})^{-2\Delta_\sigma} \) by setting \( \Delta_\sigma = 0.60 \)(the finite size extrapolated value). and four-point correlators (b)$f_{\sigma\sigma\sigma\sigma}(r=1,\theta)$ (c)$f_{\epsilon\sigma\sigma\epsilon}(r=1,\theta)$,(d)$f_{T_{\mu\nu}\sigma\sigma T_{\rho\eta}}(r=1,\theta)$ for different system sizes.}
    \label{fig:correlators}
\end{figure}

In the following, we use the \(\sigma\) operator as an example to demonstrate how to study the correlation functions of operators using the fuzzy sphere. The decomposition of the \(\hat{n}_z(\Omega)\) operator is as follows: 
\begin{equation}
\begin{aligned}
    \hat{n}_z(\Omega)=&[c_\sigma \hat{\sigma}(\Omega)+\cdots]+[c_{\sigma'} \hat{\sigma'}(\Omega)+\cdots]
    \\
    &+[c_{\sigma^{\prime\prime}} \hat{\sigma^{\prime\prime}}(\Omega)+\cdots]+\cdots
    \end{aligned}
    \label{eq:nz_decomposition}
\end{equation}
Here, ``···" represents the descendant fields corresponding to each primary operator. As already explained in \cite{Four_Han2023,OPE_hu2023}, the leading order of the correlation function of \(\hat{n}_z(\Omega)\) can provide the correlators of \(\sigma\).
\begin{equation}
\begin{aligned}
        &f_{\phi_1\sigma\sigma\phi_4}(r=1,\theta)=\frac{\left<\phi_4\mid \hat{n}_z(\Omega)\hat{n}_z^\dagger(\Omega)\mid\phi_1\right>}{|\left<0\mid \hat{n}_z(\Omega)\mid \sigma\right>|^2}\\
        =&\frac{\sum_{l,m}\Bar{Y}_{l,m}(\theta,\phi){Y}_{l,m}(\theta,\phi)\left<\phi_4\mid \hat{n}_{l,m}^z \hat{n}_{l,m}^{z\dagger}\mid \phi_1\right>}{\left|\sum_{l,m}Y_{l,m}(\theta,\phi)\left<0\mid \hat{n}_{l,m}^z\mid\sigma\right>\right|^2}\\
        =&\frac{\sum_{l=0}^{2s}\Bar{Y}_{l,m=0}(\theta,0){Y}_{l,m=0}(\theta,0)\left<\phi_4\mid \hat{n}_{l,m=0}^z \hat{n}_{l,m=0}^{z\dagger}\mid \phi_1\right>}{\left|\frac{1}{\sqrt{4\pi}}\left<0\mid \hat{n}_{l,m=0}^z\mid\sigma\right>\right|^2}
\end{aligned}
\end{equation}
For the two-point correlation function $f_{\sigma\sigma}$, we only need to take \(\left|\phi_1\right>\) and \(\left|\phi_4\right>\) as the ground state $\left|0\right>$. Fig.\ref{fig:correlators}(a) shows the dependence of the calculated two-point correlation function on \(\theta\). We also compare the finite-size results with an exact form $f_{\sigma\sigma}=(2\sin{\theta/2})^{-2\Delta_\sigma}+\mathcal{O}(R^{-1})$, as indicated by the dashed line in the Fig.\ref{fig:correlators}(a).

As for the four-point correlation function, \(\left|\phi_1\right>\) and \(\left|\phi_4\right>\) need to be set as the excited states corresponding to the CFT operators. The Fig.\ref{fig:correlators} shows the behavior of $f_{\sigma\sigma\sigma\sigma}$, $f_{\epsilon\sigma\sigma\epsilon}$, $f_{T_{\mu\nu}\sigma\sigma T_{\rho\eta}}$ within subfigure (b-d). As the system size increases, these correlation functions gradually converge to the same value.

Similarly, extracting the OPE coefficients also involves the inner product between the local operator and the excited states corresponding to the CFT operators, followed by integrating out the angular dependence. For instance, by applying the decomposition of the \( \hat{n}_z(\Omega) \) operator (Eq.\ref{eq:nz_decomposition}), a specific OPE coefficient \( f_{\phi_1\sigma\phi_3} \) can be calculated using the following expression:
\begin{equation}
\begin{aligned}
    f_{\phi_3\sigma\phi_1} & =\frac{\left<\phi_3\mid\hat{n}_z(\Omega)\mid\phi_1\right>}{\left<\sigma\mid\hat{n}_z(\Omega)\mid 0\right>}+\mathcal{O}(R^{-2})
    \\
    &=\frac{\left<\phi_3\mid\hat{n}_{0,0}^z\mid\phi_1\right>}{\left<\sigma\mid\hat{n}_{0,0}^z\mid 0\right>}+\mathcal{O}(R^{-2})
    \end{aligned}
\end{equation}

Table. \ref{tab:ope} presents a series of OPE coefficients evaluated by calculating wavefunction overlaps. These coefficients were linearly extrapolated from finite-size results. The OPE coefficients calculated using the \( \hat{n}_x(\Omega) \) operator and the \( \hat{O}_\epsilon(\Omega) \) operator are essentially consistent.


\begin{table}
\caption{\label{tab:ope} 
OPE coefficients extracted from the wave function overlap within the fuzzy sphere scheme.}
\begin{tabular}{c|c|c|c}
\hline \hline
$\text{OPE}$ &\text{Operator} &\text{Local operator} &\text{numerical calculation} \\ \hline

 $f_{\sigma \sigma \epsilon}$ &\multirow{3}{*}{$\sigma$} &\multirow{3}{*}{$\hat{n}_z(\Omega)$} &$1.1182$  \\

 $f_{\sigma' \sigma \epsilon}$ & & &$0.7395$ \\

 $f_{\sigma' \sigma \epsilon'}$ & & &$0.8723$ \\ \hline

  \multirow{2}{*}{$f_{\epsilon \epsilon \epsilon'}$} &\multirow{10}{*}{$\epsilon$} &$\hat{n}_x(\Omega)$ &$1.6005$  \\
  & &$\hat{\mathcal{O}}_\epsilon(\Omega)$ &$1.7138$  \\ 
\multirow{2}{*}{$f_{\sigma' \epsilon \sigma'}$} & &$\hat{n}_x(\Omega)$ &$1.3548$  \\
 & &$\hat{\mathcal{O}}_\epsilon(\Omega)$ &$1.3343$  \\ 
 \multirow{2}{*}{$f_{\epsilon \epsilon \epsilon}$} & &$\hat{n}_x(\Omega)$ &$1.8881$  \\
 & &$\hat{\mathcal{O}}_\epsilon(\Omega)$ &$1.9457$  \\ 
\multirow{2}{*}{$f_{\sigma \epsilon \sigma}$} & &$\hat{n}_x(\Omega)$ & $1.2410$ \\
 & &$\hat{\mathcal{O}}_\epsilon(\Omega)$ & $1.2507$ \\ 
\multirow{2}{*}{$f_{\epsilon' \epsilon \epsilon'}$} & &$\hat{n}_x(\Omega)$ &$2.1286$ \\
 & &$\hat{\mathcal{O}}_\epsilon(\Omega)$ &$2.1224$ \\
\hline \hline
\end{tabular}
\end{table}

\section{Discussion}
In the fuzzy sphere simulation, we identify two key observations:  1) the operator spectrum demonstrates approximate conformal symmetry along the phase boundary; 2) the conformal data drifts with varying interaction parameters and total system size.  
Next, we discuss the plausible physical origin of these observations.

Based on the above facts, a putative phase diagram is proposed in Fig. \ref{fig:3Dpotts_diagram}. The $V_1-h$ parameter plane comprises a phase transition line (red dashed line), where the current numerical computation points to first-order phase transitions, compatible with previous literature.  
Interestingly, a fixed point lies just outside the $V_1-h$ parameter space. The fixed point (blue star dot) can be reached out of the parameter plane by tuning an additional parameter $g$ ($g$ does not appear in the current microscopic model).  

What is the nature of this fixed point?
One possibility is that a unitary CFT controls it. If true, this CFT might coincide with a tricritical fixed point in the parameter space. A complete identification of this 3D CFT with $S_3$ global symmetry is necessary to pin down this picture. 
Future work may explore this direction. 
On the other hand, such a fixed point outside the model parameter space may be realized by tuning a complex coupling \cite{Gorbenko2018a,Gorbenko2018b}.

\subsection{Complex CFT scenario}
\begin{figure}
    \centering
    \includegraphics[width=0.6\linewidth]{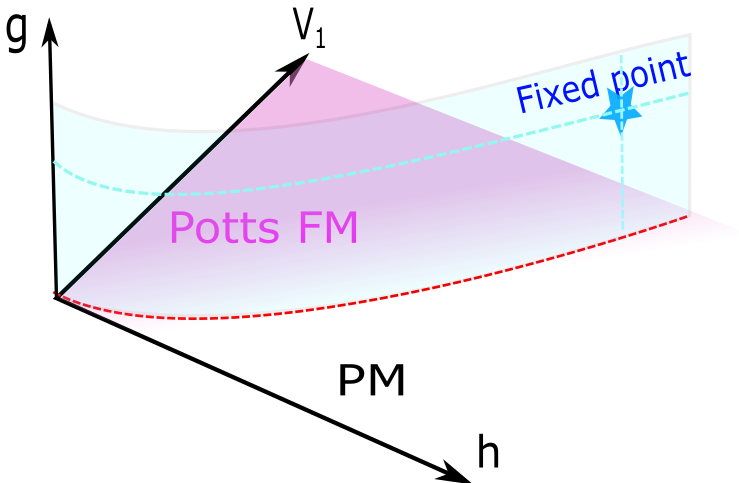}
    \caption{ The putative phase diagram is presented schematically for the 3D 3-state Potts model. 
    A phase transition line (dashed red line) appears in the $V_1-h$ parameter plane. Most of the existing numerical results support a first-order type transition. The exact fixed point (blue star dot) may exist and potentially be reached by tuning additional parameter $g$ (which is not included in the current model)  \cite{Gorbenko2018a}.
    }
    \label{fig:3Dpotts_diagram}
\end{figure}

Within this scenario, the parameter-$g$ in Fig. \ref{fig:3Dpotts_diagram} corresponds to some additional non-Hermitian interaction terms, and thus the first-order transition line is governed by conformal fixed point living in the complex parameter space. 

Complex CFT emerged initially from investigations on conformal windows of 4d gauge theories \cite{Kaplan2009}
and subsequently broadened its scope to account for some weakly first-order phase transitions in statistical physics \cite{Gorbenko2018a}. Considering a theory space spanned by some underlying degrees of freedom  (e.g. numbers of fermion flavors $N_f$ for QCD or local spin components $Q$ for Potts model), which possess an ultraviolet (UV) multicritical fixed point and an infrared (IR) stable critical point within certain regimes. Tuning the theory degrees of freedom (such as $N_f$ or $Q$), these two fixed points approach each other,
until they ultimately merge into a single conformal fixed point. Beyond this range, the fixed points might disappear within the real (unitary) theory space, while novel fixed points emerge within the realm of complex (non-unitary) theory space. These conjugated fixed points are described by complex CFTs \cite{Gorbenko2018a}, characterized by complex scaling dimensions and complex OPE coefficients. When the theory passes between these complex fixed points along the real axis, the system exhibits exponentially slow walking RG flows \cite{Gorbenko2018a}, displaying approximate scaling invariance across a large energy scale. In the Potts model context, the continuous phase transitions turn into weakly first-order type when the number of spin components $Q$ exceeds a critical value $Q>Q_c(D)$. This possibility might account for the approximate conformal symmetry and slow drifting of scaling dimensions observed in our numerical computation.


Specifically, we could describe the effective field theory of our model by $H(V_0,V_1,h) \approx H_{0}+g_{\epsilon} \int \mathrm{d}\Omega \cdot \hat{\epsilon}+g_{\epsilon'}\int \mathrm{d} \Omega \cdot \hat{\epsilon}'+(\text{irrelevant perturbation})$, where the 3D $S_3$ complex fixed points might locate at $g_{\epsilon}=0$ with $g_{\epsilon'}=\pm i \sqrt{a(Q-Q_c(D))}$ (if we assume the RG equation takes the same form as 2D cases \cite{Delfino2000}). When we tune the parameter $V_1/V_0$ and $h$ of our lattice model within the real space, $g_{\epsilon}$ and $g_{\epsilon'}$ can only take real values. For any given $V_1/V_0$, tuning $h$ changes the values of $g_{\epsilon}$ and $g_{\epsilon'}$ simultaneously. At some point $h=h_c(V_0,V_1)$, the coupling of the strongly relevant operator $g_{\epsilon}$ becomes vanishingly small. In contrast, the other coupling $g_{\epsilon'}$ remains real and finite (at least deviates from the exact fixed points with a non-vanishing imaginary part). This point corresponds to the order-disorder phase transition point shown in Fig.~\ref{fig:phasediagram}, and it always exhibits approximate conformal symmetry due to the nearby complex fixed points. However, this complex fixed point could never be reached within Hermitian Hamiltonian space, resulting in the first-order transition observed in previous literature and this work. Theoretically, one could add $S_3$ singlet operator and examine the whole phase diagram within non-Hermitian parameter space to find the exact fixed points.

In recent work, the 3-state critical and tricritical Potts model was numerically bootstrapped through the navigator function \cite{chester2022}. With the increase of spacetime dimensions, the scaling dimensions for the lowest four operators $\sigma,\sigma',\epsilon,\epsilon'$ (the last operator only for the tricritical branch) were found to be closer and closer, coinciding around a critical dimension $D_c \approx 2.6$. This calculation partially supports the above picture.

At last, we have to mention that the challenging point is to directly locate the complex fixed point in the putative phase diagram Fig. \ref{fig:3Dpotts_diagram}. This requires studying a non-Hermitian Potts model and identifying the critical points there. Future work may explore this direction to pin down the complex CFT scenario in the 3D Potts model.

\subsection{Finite-size effect}
The finite-size effect inevitably exists since our current calculations are limited to a finite number of (up to 14) quantum Potts spins. 
Our general scheme is to determine the phase transition points using the traditional order parameter, and then inspect the operator spectrum by setting on the identified transition points.
Generally, several sources of the finite-size effect may influence our conclusion. First, the finite-size scaling of rescaled order parameters (e.g., Fig. \ref{fig:eta_extrapolate}) and binder ratios (e.g., Fig. \ref{fig:U4}) may suffer from a strong finite-size effect, leading to the inaccuracy of the location of transition point $h_c$. Second, since the primaries flow with the system size $N_o$ (e.g. Fig. \ref{fig:scaling}), it is difficult to estimate accurate conformal data in the thermodynamic limit. 
The finite-size effect is not so evident for other primaries, except for $\epsilon'$. We hope the future work could resolve the scaling dimension of $\epsilon'$.

\section{Summary}
In this work, we constructed a model on the fuzzy sphere to study a phase transition belonging to the 3-dimensional 3-state Potts universal class. 
Crucially, we observed that the energy spectrum exhibits a nearly integer-spaced tower structure for the primary operators and their descendants, which unveils an approximate conformal symmetry emergent at the phase transition. We further examined the drifting behavior of the scaling dimensions for several primary operators.
This indicates that the phase transition is governed by some underlying conformal field theory, while its intrinsic nature is still unclear.
Nevertheless, unraveling the hidden conformality provides key insights into understanding the nature of phase transition in the 3D Potts model.

We envision this work could stimulate future study in several directions. 
Firstly, when we discuss the subleading singlet operator $\epsilon'$, we are less sure about its scaling dimension, due to the significant finite-size effect. We anticipate that future research will be performed within larger system sizes to determine the relevance of this primary with greater certainty.
Secondly, there is a strong desire to explore numerical bootstrap calculations for the 3D 3-state Potts universality class and verify whether the conformal data presented in this work falls within the permissible range. The data provided in this paper should serve as essential inputs for such bootstrap calculations.
Thirdly, we anticipate the simulation of the lattice model may answer additional questions proposed in this work. For example, one could study the three-dimensional 3-state Potts lattice model with $J_1$ ferromagnetic couplings together with $J_2$ antiferromagnetic couplings\cite{PhysRevB.46.944,BROWN1989412}. Although previous literature has confirmed a first-order transition in the $J_1$ only Potts model, the possibility of a second-order phase transition by adding more competing interactions is not excluded.


\begin{acknowledgments}  
W.Z. thanks Yin-Chen He for collaboration on the related projects. We thank Liangdong Hu, Youjin Deng, Yang Qi, Guangming Zhang, and Zheng Zhou for fruitful discussions. This work is supported by the National Key Research and Development Program of China Grant No. 2022YFA1402204 and the National Natural Science Foundation of China Grant Nos. 12474144, 12274086.
\end{acknowledgments}

	
	


\bibliography{Z3Sphere.bib}

\begin{thebibliography}{55}%
\makeatletter
\providecommand \@ifxundefined [1]{%
 \@ifx{#1\undefined}
}%
\providecommand \@ifnum [1]{%
 \ifnum #1\expandafter \@firstoftwo
 \else \expandafter \@secondoftwo
 \fi
}%
\providecommand \@ifx [1]{%
 \ifx #1\expandafter \@firstoftwo
 \else \expandafter \@secondoftwo
 \fi
}%
\providecommand \natexlab [1]{#1}%
\providecommand \enquote  [1]{``#1''}%
\providecommand \bibnamefont  [1]{#1}%
\providecommand \bibfnamefont [1]{#1}%
\providecommand \citenamefont [1]{#1}%
\providecommand \href@noop [0]{\@secondoftwo}%
\providecommand \href [0]{\begingroup \@sanitize@url \@href}%
\providecommand \@href[1]{\@@startlink{#1}\@@href}%
\providecommand \@@href[1]{\endgroup#1\@@endlink}%
\providecommand \@sanitize@url [0]{\catcode `\\12\catcode `\$12\catcode `\&12\catcode `\#12\catcode `\^12\catcode `\_12\catcode `\%12\relax}%
\providecommand \@@startlink[1]{}%
\providecommand \@@endlink[0]{}%
\providecommand \url  [0]{\begingroup\@sanitize@url \@url }%
\providecommand \@url [1]{\endgroup\@href {#1}{\urlprefix }}%
\providecommand \urlprefix  [0]{URL }%
\providecommand \Eprint [0]{\href }%
\providecommand \doibase [0]{https://doi.org/}%
\providecommand \selectlanguage [0]{\@gobble}%
\providecommand \bibinfo  [0]{\@secondoftwo}%
\providecommand \bibfield  [0]{\@secondoftwo}%
\providecommand \translation [1]{[#1]}%
\providecommand \BibitemOpen [0]{}%
\providecommand \bibitemStop [0]{}%
\providecommand \bibitemNoStop [0]{.\EOS\space}%
\providecommand \EOS [0]{\spacefactor3000\relax}%
\providecommand \BibitemShut  [1]{\csname bibitem#1\endcsname}%
\let\auto@bib@innerbib\@empty
\bibitem [{\citenamefont {Polyakov}(1970)}]{polyakov1970conformal}%
  \BibitemOpen
  \bibfield  {author} {\bibinfo {author} {\bibfnamefont {A.~M.}\ \bibnamefont {Polyakov}},\ }\bibfield  {title} {\bibinfo {title} {Conformal symmetry of critical fluctuations},\ }\href@noop {} {\bibfield  {journal} {\bibinfo  {journal} {JETP Lett.}\ }\textbf {\bibinfo {volume} {12}},\ \bibinfo {pages} {381} (\bibinfo {year} {1970})}\BibitemShut {NoStop}%
\bibitem [{\citenamefont {Cardy}(1996)}]{Cardy_book}%
  \BibitemOpen
  \bibfield  {author} {\bibinfo {author} {\bibfnamefont {J.}~\bibnamefont {Cardy}},\ }\href@noop {} {\emph {\bibinfo {title} {Scaling and Renormalization in Statistical Physics}}}\ (\bibinfo  {publisher} {Cambridge University Press, Cambridge, England},\ \bibinfo {year} {1996})\BibitemShut {NoStop}%
\bibitem [{\citenamefont {Henkel}(1999)}]{Henkel_book}%
  \BibitemOpen
  \bibfield  {author} {\bibinfo {author} {\bibfnamefont {M.}~\bibnamefont {Henkel}},\ }\href@noop {} {\emph {\bibinfo {title} {Conformal invariance and critical phenomena}}}\ (\bibinfo  {publisher} {Springer Science \& Business Media},\ \bibinfo {year} {1999})\BibitemShut {NoStop}%
\bibitem [{\citenamefont {Philippe~Francesco}(1997)}]{yellowbook}%
  \BibitemOpen
  \bibfield  {author} {\bibinfo {author} {\bibfnamefont {D.~S.}\ \bibnamefont {Philippe~Francesco}, \bibfnamefont {Pierre~Mathieu}},\ }\href@noop {} {\emph {\bibinfo {title} {Conformal Field Theory}}},\ Graduate Texts in Contemporary Physics\ (\bibinfo  {publisher} {Springer New York, NY},\ \bibinfo {year} {1997})\BibitemShut {NoStop}%
\bibitem [{\citenamefont {Gorbenko}\ \emph {et~al.}(2018{\natexlab{a}})\citenamefont {Gorbenko}, \citenamefont {Rychkov},\ and\ \citenamefont {Zan}}]{Gorbenko2018a}%
  \BibitemOpen
  \bibfield  {author} {\bibinfo {author} {\bibfnamefont {V.}~\bibnamefont {Gorbenko}}, \bibinfo {author} {\bibfnamefont {S.}~\bibnamefont {Rychkov}},\ and\ \bibinfo {author} {\bibfnamefont {B.}~\bibnamefont {Zan}},\ }\bibfield  {title} {\bibinfo {title} {Walking, weak first-order transitions, and complex cfts},\ }\bibfield  {journal} {\bibinfo  {journal} {Journal of High Energy Physics}\ }\textbf {\bibinfo {volume} {2018}},\ \href {https://doi.org/10.1007/jhep10(2018)108} {10.1007/jhep10(2018)108} (\bibinfo {year} {2018}{\natexlab{a}})\BibitemShut {NoStop}%
\bibitem [{\citenamefont {Gorbenko}\ \emph {et~al.}(2018{\natexlab{b}})\citenamefont {Gorbenko}, \citenamefont {Rychkov},\ and\ \citenamefont {Zan}}]{Gorbenko2018b}%
  \BibitemOpen
  \bibfield  {author} {\bibinfo {author} {\bibfnamefont {V.}~\bibnamefont {Gorbenko}}, \bibinfo {author} {\bibfnamefont {S.}~\bibnamefont {Rychkov}},\ and\ \bibinfo {author} {\bibfnamefont {B.}~\bibnamefont {Zan}},\ }\bibfield  {title} {\bibinfo {title} {{Walking, Weak first-order transitions, and Complex CFTs II. Two-dimensional Potts model at $Q>4$}},\ }\href {https://doi.org/10.21468/SciPostPhys.5.5.050} {\bibfield  {journal} {\bibinfo  {journal} {SciPost Phys.}\ }\textbf {\bibinfo {volume} {5}},\ \bibinfo {pages} {050} (\bibinfo {year} {2018}{\natexlab{b}})}\BibitemShut {NoStop}%
\bibitem [{\citenamefont {Kaplan}\ \emph {et~al.}(2009)\citenamefont {Kaplan}, \citenamefont {Lee}, \citenamefont {Son},\ and\ \citenamefont {Stephanov}}]{Kaplan2009}%
  \BibitemOpen
  \bibfield  {author} {\bibinfo {author} {\bibfnamefont {D.~B.}\ \bibnamefont {Kaplan}}, \bibinfo {author} {\bibfnamefont {J.-W.}\ \bibnamefont {Lee}}, \bibinfo {author} {\bibfnamefont {D.~T.}\ \bibnamefont {Son}},\ and\ \bibinfo {author} {\bibfnamefont {M.~A.}\ \bibnamefont {Stephanov}},\ }\bibfield  {title} {\bibinfo {title} {Conformality lost},\ }\href {https://doi.org/10.1103/PhysRevD.80.125005} {\bibfield  {journal} {\bibinfo  {journal} {Phys. Rev. D}\ }\textbf {\bibinfo {volume} {80}},\ \bibinfo {pages} {125005} (\bibinfo {year} {2009})}\BibitemShut {NoStop}%
\bibitem [{\citenamefont {Jacobsen}\ and\ \citenamefont {Wiese}(2024)}]{jacobsen2024lattice}%
  \BibitemOpen
  \bibfield  {author} {\bibinfo {author} {\bibfnamefont {J.~L.}\ \bibnamefont {Jacobsen}}\ and\ \bibinfo {author} {\bibfnamefont {K.~J.}\ \bibnamefont {Wiese}},\ }\bibfield  {title} {\bibinfo {title} {Lattice realization of complex conformal field theories: Two-dimensional potts model with $q>4$ states},\ }\href {https://doi.org/10.1103/PhysRevLett.133.077101} {\bibfield  {journal} {\bibinfo  {journal} {Phys. Rev. Lett.}\ }\textbf {\bibinfo {volume} {133}},\ \bibinfo {pages} {077101} (\bibinfo {year} {2024})}\BibitemShut {NoStop}%
\bibitem [{\citenamefont {Tang}\ \emph {et~al.}(2024)\citenamefont {Tang}, \citenamefont {Ma}, \citenamefont {Tang}, \citenamefont {He},\ and\ \citenamefont {Zhu}}]{tang2024}%
  \BibitemOpen
  \bibfield  {author} {\bibinfo {author} {\bibfnamefont {Y.}~\bibnamefont {Tang}}, \bibinfo {author} {\bibfnamefont {H.}~\bibnamefont {Ma}}, \bibinfo {author} {\bibfnamefont {Q.}~\bibnamefont {Tang}}, \bibinfo {author} {\bibfnamefont {Y.-C.}\ \bibnamefont {He}},\ and\ \bibinfo {author} {\bibfnamefont {W.}~\bibnamefont {Zhu}},\ }\bibfield  {title} {\bibinfo {title} {Reclaiming the lost conformality in a non-hermitian quantum 5-state potts model},\ }\href {https://doi.org/10.1103/PhysRevLett.133.076504} {\bibfield  {journal} {\bibinfo  {journal} {Phys. Rev. Lett.}\ }\textbf {\bibinfo {volume} {133}},\ \bibinfo {pages} {076504} (\bibinfo {year} {2024})}\BibitemShut {NoStop}%
\bibitem [{\citenamefont {Zhu}\ \emph {et~al.}(2023)\citenamefont {Zhu}, \citenamefont {Han}, \citenamefont {Huffman}, \citenamefont {Hofmann},\ and\ \citenamefont {He}}]{ZHHHH2022}%
  \BibitemOpen
  \bibfield  {author} {\bibinfo {author} {\bibfnamefont {W.}~\bibnamefont {Zhu}}, \bibinfo {author} {\bibfnamefont {C.}~\bibnamefont {Han}}, \bibinfo {author} {\bibfnamefont {E.}~\bibnamefont {Huffman}}, \bibinfo {author} {\bibfnamefont {J.~S.}\ \bibnamefont {Hofmann}},\ and\ \bibinfo {author} {\bibfnamefont {Y.-C.}\ \bibnamefont {He}},\ }\bibfield  {title} {\bibinfo {title} {Uncovering conformal symmetry in the 3d ising transition: State-operator correspondence from a quantum fuzzy sphere regularization},\ }\href {https://doi.org/10.1103/PhysRevX.13.021009} {\bibfield  {journal} {\bibinfo  {journal} {Phys. Rev. X}\ }\textbf {\bibinfo {volume} {13}},\ \bibinfo {pages} {021009} (\bibinfo {year} {2023})}\BibitemShut {NoStop}%
\bibitem [{\citenamefont {Hu}\ \emph {et~al.}(2023)\citenamefont {Hu}, \citenamefont {He},\ and\ \citenamefont {Zhu}}]{OPE_hu2023}%
  \BibitemOpen
  \bibfield  {author} {\bibinfo {author} {\bibfnamefont {L.}~\bibnamefont {Hu}}, \bibinfo {author} {\bibfnamefont {Y.-C.}\ \bibnamefont {He}},\ and\ \bibinfo {author} {\bibfnamefont {W.}~\bibnamefont {Zhu}},\ }\bibfield  {title} {\bibinfo {title} {Operator product expansion coefficients of the 3d ising criticality via quantum fuzzy spheres},\ }\href {https://doi.org/10.1103/PhysRevLett.131.031601} {\bibfield  {journal} {\bibinfo  {journal} {Phys. Rev. Lett.}\ }\textbf {\bibinfo {volume} {131}},\ \bibinfo {pages} {031601} (\bibinfo {year} {2023})}\BibitemShut {NoStop}%
\bibitem [{\citenamefont {Han}\ \emph {et~al.}(2023{\natexlab{a}})\citenamefont {Han}, \citenamefont {Hu},\ and\ \citenamefont {Zhu}}]{han2023conformaloperatorcontentwilsonfisher}%
  \BibitemOpen
  \bibfield  {author} {\bibinfo {author} {\bibfnamefont {C.}~\bibnamefont {Han}}, \bibinfo {author} {\bibfnamefont {L.}~\bibnamefont {Hu}},\ and\ \bibinfo {author} {\bibfnamefont {W.}~\bibnamefont {Zhu}},\ }\href {https://arxiv.org/abs/2312.04047} {\bibinfo {title} {Conformal operator content of the wilson-fisher transition on fuzzy sphere bilayers}} (\bibinfo {year} {2023}{\natexlab{a}}),\ \Eprint {https://arxiv.org/abs/2312.04047} {arXiv:2312.04047 [cond-mat.str-el]} \BibitemShut {NoStop}%
\bibitem [{\citenamefont {Zhou}\ \emph {et~al.}(2024)\citenamefont {Zhou}, \citenamefont {Hu}, \citenamefont {Zhu},\ and\ \citenamefont {He}}]{zhou2024mathrmso5deconfinedphasetransition}%
  \BibitemOpen
  \bibfield  {author} {\bibinfo {author} {\bibfnamefont {Z.}~\bibnamefont {Zhou}}, \bibinfo {author} {\bibfnamefont {L.}~\bibnamefont {Hu}}, \bibinfo {author} {\bibfnamefont {W.}~\bibnamefont {Zhu}},\ and\ \bibinfo {author} {\bibfnamefont {Y.-C.}\ \bibnamefont {He}},\ }\bibfield  {title} {\bibinfo {title} {So(5) deconfined phase transition under the fuzzy-sphere microscope: Approximate conformal symmetry, pseudo-criticality, and operator spectrum},\ }\href {https://doi.org/10.1103/PhysRevX.14.021044} {\bibfield  {journal} {\bibinfo  {journal} {Phys. Rev. X}\ }\textbf {\bibinfo {volume} {14}},\ \bibinfo {pages} {021044} (\bibinfo {year} {2024})}\BibitemShut {NoStop}%
\bibitem [{\citenamefont {Wu}(1982)}]{RevModPhys.54.235}%
  \BibitemOpen
  \bibfield  {author} {\bibinfo {author} {\bibfnamefont {F.~Y.}\ \bibnamefont {Wu}},\ }\bibfield  {title} {\bibinfo {title} {The potts model},\ }\href {https://doi.org/10.1103/RevModPhys.54.235} {\bibfield  {journal} {\bibinfo  {journal} {Rev. Mod. Phys.}\ }\textbf {\bibinfo {volume} {54}},\ \bibinfo {pages} {235} (\bibinfo {year} {1982})}\BibitemShut {NoStop}%
\bibitem [{\citenamefont {Potts}(1952)}]{Potts1952}%
  \BibitemOpen
  \bibfield  {author} {\bibinfo {author} {\bibfnamefont {R.~B.}\ \bibnamefont {Potts}},\ }\bibfield  {title} {\bibinfo {title} {Some generalized order-disorder transformations},\ }\href {https://doi.org/10.1017/S0305004100027419} {\bibfield  {journal} {\bibinfo  {journal} {Mathematical Proceedings of the Cambridge Philosophical Society}\ }\textbf {\bibinfo {volume} {48}},\ \bibinfo {pages} {106–109} (\bibinfo {year} {1952})}\BibitemShut {NoStop}%
\bibitem [{\citenamefont {Fishman}\ \emph {et~al.}(2022)\citenamefont {Fishman}, \citenamefont {White},\ and\ \citenamefont {Stoudenmire}}]{itensor}%
  \BibitemOpen
  \bibfield  {author} {\bibinfo {author} {\bibfnamefont {M.}~\bibnamefont {Fishman}}, \bibinfo {author} {\bibfnamefont {S.~R.}\ \bibnamefont {White}},\ and\ \bibinfo {author} {\bibfnamefont {E.~M.}\ \bibnamefont {Stoudenmire}},\ }\bibfield  {title} {\bibinfo {title} {{The ITensor Software Library for Tensor Network Calculations}},\ }\href {https://doi.org/10.21468/SciPostPhysCodeb.4} {\bibfield  {journal} {\bibinfo  {journal} {SciPost Phys. Codebases}\ ,\ \bibinfo {pages} {4}} (\bibinfo {year} {2022})}\BibitemShut {NoStop}%
\bibitem [{\citenamefont {Gennes}(1971)}]{doi:10.1080/15421407108082773}%
  \BibitemOpen
  \bibfield  {author} {\bibinfo {author} {\bibfnamefont {P.~G.~D.}\ \bibnamefont {Gennes}},\ }\bibfield  {title} {\bibinfo {title} {Short range order effects in the isotropic phase of nematics and cholesterics},\ }\href {https://doi.org/10.1080/15421407108082773} {\bibfield  {journal} {\bibinfo  {journal} {Molecular Crystals and Liquid Crystals}\ }\textbf {\bibinfo {volume} {12}},\ \bibinfo {pages} {193} (\bibinfo {year} {1971})},\ \Eprint {https://arxiv.org/abs/https://doi.org/10.1080/15421407108082773} {https://doi.org/10.1080/15421407108082773} \BibitemShut {NoStop}%
\bibitem [{\citenamefont {Mukamel}\ \emph {et~al.}(1976)\citenamefont {Mukamel}, \citenamefont {Fisher},\ and\ \citenamefont {Domany}}]{PhysRevLett.37.565}%
  \BibitemOpen
  \bibfield  {author} {\bibinfo {author} {\bibfnamefont {D.}~\bibnamefont {Mukamel}}, \bibinfo {author} {\bibfnamefont {M.~E.}\ \bibnamefont {Fisher}},\ and\ \bibinfo {author} {\bibfnamefont {E.}~\bibnamefont {Domany}},\ }\bibfield  {title} {\bibinfo {title} {Magnetization of cubic ferromagnets and the three-component potts model},\ }\href {https://doi.org/10.1103/PhysRevLett.37.565} {\bibfield  {journal} {\bibinfo  {journal} {Phys. Rev. Lett.}\ }\textbf {\bibinfo {volume} {37}},\ \bibinfo {pages} {565} (\bibinfo {year} {1976})}\BibitemShut {NoStop}%
\bibitem [{\citenamefont {Weger}\ and\ \citenamefont {Goldberg}(1974)}]{WEGER19741}%
  \BibitemOpen
  \bibfield  {author} {\bibinfo {author} {\bibfnamefont {M.}~\bibnamefont {Weger}}\ and\ \bibinfo {author} {\bibfnamefont {I.}~\bibnamefont {Goldberg}},\ }\bibfield  {title} {\bibinfo {title} {Some lattice and electronic properties of the beta-tungstens}\ }(\bibinfo  {publisher} {Academic Press},\ \bibinfo {year} {1974})\ pp.\ \bibinfo {pages} {1--177}\BibitemShut {NoStop}%
\bibitem [{\citenamefont {Cardy}\ \emph {et~al.}(1980)\citenamefont {Cardy}, \citenamefont {Nauenberg},\ and\ \citenamefont {Scalapino}}]{Cardy1980}%
  \BibitemOpen
  \bibfield  {author} {\bibinfo {author} {\bibfnamefont {J.~L.}\ \bibnamefont {Cardy}}, \bibinfo {author} {\bibfnamefont {M.}~\bibnamefont {Nauenberg}},\ and\ \bibinfo {author} {\bibfnamefont {D.~J.}\ \bibnamefont {Scalapino}},\ }\bibfield  {title} {\bibinfo {title} {Scaling theory of the potts-model multicritical point},\ }\href {https://doi.org/10.1103/PhysRevB.22.2560} {\bibfield  {journal} {\bibinfo  {journal} {Phys. Rev. B}\ }\textbf {\bibinfo {volume} {22}},\ \bibinfo {pages} {2560} (\bibinfo {year} {1980})}\BibitemShut {NoStop}%
\bibitem [{\citenamefont {Nauenberg}\ and\ \citenamefont {Scalapino}(1980)}]{Nauenberg1980}%
  \BibitemOpen
  \bibfield  {author} {\bibinfo {author} {\bibfnamefont {M.}~\bibnamefont {Nauenberg}}\ and\ \bibinfo {author} {\bibfnamefont {D.~J.}\ \bibnamefont {Scalapino}},\ }\bibfield  {title} {\bibinfo {title} {Singularities and scaling functions at the potts-model multicritical point},\ }\href {https://doi.org/10.1103/PhysRevLett.44.837} {\bibfield  {journal} {\bibinfo  {journal} {Phys. Rev. Lett.}\ }\textbf {\bibinfo {volume} {44}},\ \bibinfo {pages} {837} (\bibinfo {year} {1980})}\BibitemShut {NoStop}%
\bibitem [{\citenamefont {S\'olyom}\ and\ \citenamefont {Pfeuty}(1981)}]{Pfeuty1981}%
  \BibitemOpen
  \bibfield  {author} {\bibinfo {author} {\bibfnamefont {J.}~\bibnamefont {S\'olyom}}\ and\ \bibinfo {author} {\bibfnamefont {P.}~\bibnamefont {Pfeuty}},\ }\bibfield  {title} {\bibinfo {title} {Renormalization-group study of the hamiltonian version of the potts model},\ }\href {https://doi.org/10.1103/PhysRevB.24.218} {\bibfield  {journal} {\bibinfo  {journal} {Phys. Rev. B}\ }\textbf {\bibinfo {volume} {24}},\ \bibinfo {pages} {218} (\bibinfo {year} {1981})}\BibitemShut {NoStop}%
\bibitem [{\citenamefont {Newman}\ \emph {et~al.}(1984)\citenamefont {Newman}, \citenamefont {Riedel},\ and\ \citenamefont {Muto}}]{Newman1984}%
  \BibitemOpen
  \bibfield  {author} {\bibinfo {author} {\bibfnamefont {K.~E.}\ \bibnamefont {Newman}}, \bibinfo {author} {\bibfnamefont {E.~K.}\ \bibnamefont {Riedel}},\ and\ \bibinfo {author} {\bibfnamefont {S.}~\bibnamefont {Muto}},\ }\bibfield  {title} {\bibinfo {title} {$q$-state potts model by wilson's exact renormalization-group equation},\ }\href {https://doi.org/10.1103/PhysRevB.29.302} {\bibfield  {journal} {\bibinfo  {journal} {Phys. Rev. B}\ }\textbf {\bibinfo {volume} {29}},\ \bibinfo {pages} {302} (\bibinfo {year} {1984})}\BibitemShut {NoStop}%
\bibitem [{\citenamefont {Baxter}(1973)}]{Baxter1973}%
  \BibitemOpen
  \bibfield  {author} {\bibinfo {author} {\bibfnamefont {R.~J.}\ \bibnamefont {Baxter}},\ }\bibfield  {title} {\bibinfo {title} {Potts model at the critical temperature},\ }\href@noop {} {\bibfield  {journal} {\bibinfo  {journal} {Journal of Physics C: Solid State Physics}\ }\textbf {\bibinfo {volume} {6}},\ \bibinfo {pages} {L445} (\bibinfo {year} {1973})}\BibitemShut {NoStop}%
\bibitem [{\citenamefont {Nienhuis}\ \emph {et~al.}(1980)\citenamefont {Nienhuis}, \citenamefont {Riedel},\ and\ \citenamefont {Schick}}]{Nienhuis1980}%
  \BibitemOpen
  \bibfield  {author} {\bibinfo {author} {\bibfnamefont {B.}~\bibnamefont {Nienhuis}}, \bibinfo {author} {\bibfnamefont {E.}~\bibnamefont {Riedel}},\ and\ \bibinfo {author} {\bibfnamefont {M.}~\bibnamefont {Schick}},\ }\bibfield  {title} {\bibinfo {title} {Variational renormalisation-group approach to the q-state potts model in two dimensions},\ }\href@noop {} {\bibfield  {journal} {\bibinfo  {journal} {Journal of Physics A: Mathematical and General}\ }\textbf {\bibinfo {volume} {13}},\ \bibinfo {pages} {L31} (\bibinfo {year} {1980})}\BibitemShut {NoStop}%
\bibitem [{\citenamefont {Buddenoir}\ and\ \citenamefont {Wallon}(1993)}]{Buddenoir1993}%
  \BibitemOpen
  \bibfield  {author} {\bibinfo {author} {\bibfnamefont {E.}~\bibnamefont {Buddenoir}}\ and\ \bibinfo {author} {\bibfnamefont {S.}~\bibnamefont {Wallon}},\ }\bibfield  {title} {\bibinfo {title} {The correlation length of the potts model at the first-order transition point},\ }\href {https://doi.org/10.1088/0305-4470/26/13/009} {\bibfield  {journal} {\bibinfo  {journal} {Journal of Physics A: Mathematical and General}\ }\textbf {\bibinfo {volume} {26}},\ \bibinfo {pages} {3045} (\bibinfo {year} {1993})}\BibitemShut {NoStop}%
\bibitem [{\citenamefont {Haldar}\ \emph {et~al.}(2023)\citenamefont {Haldar}, \citenamefont {Tavakol}, \citenamefont {Ma},\ and\ \citenamefont {Scaffidi}}]{Haldar2023}%
  \BibitemOpen
  \bibfield  {author} {\bibinfo {author} {\bibfnamefont {A.}~\bibnamefont {Haldar}}, \bibinfo {author} {\bibfnamefont {O.}~\bibnamefont {Tavakol}}, \bibinfo {author} {\bibfnamefont {H.}~\bibnamefont {Ma}},\ and\ \bibinfo {author} {\bibfnamefont {T.}~\bibnamefont {Scaffidi}},\ }\bibfield  {title} {\bibinfo {title} {Hidden critical points in the two-dimensional $o (n > 2)$ model: Exact numerical study of a complex conformal field theory},\ }\href {https://doi.org/10.1103/PhysRevLett.131.131601} {\bibfield  {journal} {\bibinfo  {journal} {Phys. Rev. Lett.}\ }\textbf {\bibinfo {volume} {131}},\ \bibinfo {pages} {131601} (\bibinfo {year} {2023})}\BibitemShut {NoStop}%
\bibitem [{\citenamefont {Nienhuis}\ \emph {et~al.}(1981)\citenamefont {Nienhuis}, \citenamefont {Riedel},\ and\ \citenamefont {Schick}}]{Nienhuis1981}%
  \BibitemOpen
  \bibfield  {author} {\bibinfo {author} {\bibfnamefont {B.}~\bibnamefont {Nienhuis}}, \bibinfo {author} {\bibfnamefont {E.~K.}\ \bibnamefont {Riedel}},\ and\ \bibinfo {author} {\bibfnamefont {M.}~\bibnamefont {Schick}},\ }\bibfield  {title} {\bibinfo {title} {$q$-state potts model in general dimension},\ }\href {https://doi.org/10.1103/PhysRevB.23.6055} {\bibfield  {journal} {\bibinfo  {journal} {Phys. Rev. B}\ }\textbf {\bibinfo {volume} {23}},\ \bibinfo {pages} {6055} (\bibinfo {year} {1981})}\BibitemShut {NoStop}%
\bibitem [{\citenamefont {S\'anchez-Villalobos}\ \emph {et~al.}(2023)\citenamefont {S\'anchez-Villalobos}, \citenamefont {Delamotte},\ and\ \citenamefont {Wschebor}}]{Villalobos2023}%
  \BibitemOpen
  \bibfield  {author} {\bibinfo {author} {\bibfnamefont {C.~A.}\ \bibnamefont {S\'anchez-Villalobos}}, \bibinfo {author} {\bibfnamefont {B.}~\bibnamefont {Delamotte}},\ and\ \bibinfo {author} {\bibfnamefont {N.}~\bibnamefont {Wschebor}},\ }\bibfield  {title} {\bibinfo {title} {$q$-state potts model from the nonperturbative renormalization group},\ }\href {https://doi.org/10.1103/PhysRevE.108.064120} {\bibfield  {journal} {\bibinfo  {journal} {Phys. Rev. E}\ }\textbf {\bibinfo {volume} {108}},\ \bibinfo {pages} {064120} (\bibinfo {year} {2023})}\BibitemShut {NoStop}%
\bibitem [{\citenamefont {Barkema}\ and\ \citenamefont {de~Boer}(1991)}]{Barkema1991}%
  \BibitemOpen
  \bibfield  {author} {\bibinfo {author} {\bibfnamefont {G.}~\bibnamefont {Barkema}}\ and\ \bibinfo {author} {\bibfnamefont {J.}~\bibnamefont {de~Boer}},\ }\bibfield  {title} {\bibinfo {title} {Numerical study of phase transitions in potts models},\ }\href {https://doi.org/10.1103/PhysRevA.44.8000} {\bibfield  {journal} {\bibinfo  {journal} {Phys. Rev. A}\ }\textbf {\bibinfo {volume} {44}},\ \bibinfo {pages} {8000} (\bibinfo {year} {1991})}\BibitemShut {NoStop}%
\bibitem [{\citenamefont {Lee}\ and\ \citenamefont {Kosterlitz}(1991)}]{Jooyoung1991}%
  \BibitemOpen
  \bibfield  {author} {\bibinfo {author} {\bibfnamefont {J.}~\bibnamefont {Lee}}\ and\ \bibinfo {author} {\bibfnamefont {J.~M.}\ \bibnamefont {Kosterlitz}},\ }\bibfield  {title} {\bibinfo {title} {Three-dimensional q-state potts model: Monte carlo study near q=3},\ }\href {https://doi.org/10.1103/PhysRevB.43.1268} {\bibfield  {journal} {\bibinfo  {journal} {Phys. Rev. B}\ }\textbf {\bibinfo {volume} {43}},\ \bibinfo {pages} {1268} (\bibinfo {year} {1991})}\BibitemShut {NoStop}%
\bibitem [{\citenamefont {Gliozzi}(2002)}]{Gliozzi2002}%
  \BibitemOpen
  \bibfield  {author} {\bibinfo {author} {\bibfnamefont {F.}~\bibnamefont {Gliozzi}},\ }\bibfield  {title} {\bibinfo {title} {Simulation of potts models with real q and no critical slowing down},\ }\href {https://doi.org/10.1103/PhysRevE.66.016115} {\bibfield  {journal} {\bibinfo  {journal} {Phys. Rev. E}\ }\textbf {\bibinfo {volume} {66}},\ \bibinfo {pages} {016115} (\bibinfo {year} {2002})}\BibitemShut {NoStop}%
\bibitem [{\citenamefont {Jensen}\ and\ \citenamefont {Mouritsen}(1979)}]{3DPotts_MC_1979}%
  \BibitemOpen
  \bibfield  {author} {\bibinfo {author} {\bibfnamefont {S.~J.~K.}\ \bibnamefont {Jensen}}\ and\ \bibinfo {author} {\bibfnamefont {O.~G.}\ \bibnamefont {Mouritsen}},\ }\bibfield  {title} {\bibinfo {title} {Is the phase transition of the three-state potts model continuous in three dimensions?},\ }\href {https://doi.org/10.1103/PhysRevLett.43.1736} {\bibfield  {journal} {\bibinfo  {journal} {Phys. Rev. Lett.}\ }\textbf {\bibinfo {volume} {43}},\ \bibinfo {pages} {1736} (\bibinfo {year} {1979})}\BibitemShut {NoStop}%
\bibitem [{\citenamefont {Bl\"ote}\ and\ \citenamefont {Swendsen}(1979)}]{3DPotts_MCRG_1979}%
  \BibitemOpen
  \bibfield  {author} {\bibinfo {author} {\bibfnamefont {H.~W.~J.}\ \bibnamefont {Bl\"ote}}\ and\ \bibinfo {author} {\bibfnamefont {R.~H.}\ \bibnamefont {Swendsen}},\ }\bibfield  {title} {\bibinfo {title} {First-order phase transitions and the three-state potts model},\ }\href {https://doi.org/10.1103/PhysRevLett.43.799} {\bibfield  {journal} {\bibinfo  {journal} {Phys. Rev. Lett.}\ }\textbf {\bibinfo {volume} {43}},\ \bibinfo {pages} {799} (\bibinfo {year} {1979})}\BibitemShut {NoStop}%
\bibitem [{\citenamefont {Alves}\ \emph {et~al.}(1991)\citenamefont {Alves}, \citenamefont {Berg},\ and\ \citenamefont {Villanova}}]{Alves1991}%
  \BibitemOpen
  \bibfield  {author} {\bibinfo {author} {\bibfnamefont {N.~A.}\ \bibnamefont {Alves}}, \bibinfo {author} {\bibfnamefont {B.~A.}\ \bibnamefont {Berg}},\ and\ \bibinfo {author} {\bibfnamefont {R.}~\bibnamefont {Villanova}},\ }\bibfield  {title} {\bibinfo {title} {Potts models: Density of states and mass gap from monte carlo calculations},\ }\href {https://doi.org/10.1103/PhysRevB.43.5846} {\bibfield  {journal} {\bibinfo  {journal} {Phys. Rev. B}\ }\textbf {\bibinfo {volume} {43}},\ \bibinfo {pages} {5846} (\bibinfo {year} {1991})}\BibitemShut {NoStop}%
\bibitem [{\citenamefont {Janke}\ and\ \citenamefont {Villanova}(1997)}]{JANKE1997}%
  \BibitemOpen
  \bibfield  {author} {\bibinfo {author} {\bibfnamefont {W.}~\bibnamefont {Janke}}\ and\ \bibinfo {author} {\bibfnamefont {R.}~\bibnamefont {Villanova}},\ }\bibfield  {title} {\bibinfo {title} {Three-dimensional 3-state potts model revisited with new techniques},\ }\href {https://doi.org/https://doi.org/10.1016/S0550-3213(96)00710-9} {\bibfield  {journal} {\bibinfo  {journal} {Nuclear Physics B}\ }\textbf {\bibinfo {volume} {489}},\ \bibinfo {pages} {679} (\bibinfo {year} {1997})}\BibitemShut {NoStop}%
\bibitem [{\citenamefont {Bazavov}\ and\ \citenamefont {Berg}(2007)}]{Berg2007}%
  \BibitemOpen
  \bibfield  {author} {\bibinfo {author} {\bibfnamefont {A.}~\bibnamefont {Bazavov}}\ and\ \bibinfo {author} {\bibfnamefont {B.~A.}\ \bibnamefont {Berg}},\ }\bibfield  {title} {\bibinfo {title} {Normalized entropy density of the 3d 3-state potts model},\ }\href {https://doi.org/10.1103/PhysRevD.75.094506} {\bibfield  {journal} {\bibinfo  {journal} {Phys. Rev. D}\ }\textbf {\bibinfo {volume} {75}},\ \bibinfo {pages} {094506} (\bibinfo {year} {2007})}\BibitemShut {NoStop}%
\bibitem [{\citenamefont {Wang}\ \emph {et~al.}(2014)\citenamefont {Wang}, \citenamefont {Xie}, \citenamefont {Chen}, \citenamefont {Normand},\ and\ \citenamefont {Xiang}}]{Wang2014}%
  \BibitemOpen
  \bibfield  {author} {\bibinfo {author} {\bibfnamefont {S.}~\bibnamefont {Wang}}, \bibinfo {author} {\bibfnamefont {Z.}~\bibnamefont {Xie}}, \bibinfo {author} {\bibfnamefont {J.}~\bibnamefont {Chen}}, \bibinfo {author} {\bibfnamefont {B.}~\bibnamefont {Normand}},\ and\ \bibinfo {author} {\bibfnamefont {T.}~\bibnamefont {Xiang}},\ }\bibfield  {title} {\bibinfo {title} {Phase transitions of ferromagnetic potts models on the simple cubic lattice},\ }\href {https://doi.org/10.1088/0256-307X/31/7/070503} {\bibfield  {journal} {\bibinfo  {journal} {Chinese Physics Letters}\ }\textbf {\bibinfo {volume} {31}},\ \bibinfo {pages} {070503} (\bibinfo {year} {2014})}\BibitemShut {NoStop}%
\bibitem [{\citenamefont {Nishino}\ \emph {et~al.}(2000)\citenamefont {Nishino}, \citenamefont {Okunishi}, \citenamefont {Hieida}, \citenamefont {Maeshima},\ and\ \citenamefont {Akutsu}}]{NISHINO2000}%
  \BibitemOpen
  \bibfield  {author} {\bibinfo {author} {\bibfnamefont {T.}~\bibnamefont {Nishino}}, \bibinfo {author} {\bibfnamefont {K.}~\bibnamefont {Okunishi}}, \bibinfo {author} {\bibfnamefont {Y.}~\bibnamefont {Hieida}}, \bibinfo {author} {\bibfnamefont {N.}~\bibnamefont {Maeshima}},\ and\ \bibinfo {author} {\bibfnamefont {Y.}~\bibnamefont {Akutsu}},\ }\bibfield  {title} {\bibinfo {title} {Self-consistent tensor product variational approximation for 3d classical models},\ }\href {https://doi.org/https://doi.org/10.1016/S0550-3213(00)00133-4} {\bibfield  {journal} {\bibinfo  {journal} {Nuclear Physics B}\ }\textbf {\bibinfo {volume} {575}},\ \bibinfo {pages} {504} (\bibinfo {year} {2000})}\BibitemShut {NoStop}%
\bibitem [{\citenamefont {Chester}\ and\ \citenamefont {Su}(2022)}]{chester2022}%
  \BibitemOpen
  \bibfield  {author} {\bibinfo {author} {\bibfnamefont {S.~M.}\ \bibnamefont {Chester}}\ and\ \bibinfo {author} {\bibfnamefont {N.}~\bibnamefont {Su}},\ }\href {https://arxiv.org/abs/2210.09091} {\bibinfo {title} {Upper critical dimension of the 3-state potts model}} (\bibinfo {year} {2022}),\ \Eprint {https://arxiv.org/abs/2210.09091} {arXiv:2210.09091 [hep-th]} \BibitemShut {NoStop}%
\bibitem [{\citenamefont {Grollau}\ \emph {et~al.}(2001)\citenamefont {Grollau}, \citenamefont {Rosinberg},\ and\ \citenamefont {Tarjus}}]{GROLLAU2001}%
  \BibitemOpen
  \bibfield  {author} {\bibinfo {author} {\bibfnamefont {S.}~\bibnamefont {Grollau}}, \bibinfo {author} {\bibfnamefont {M.}~\bibnamefont {Rosinberg}},\ and\ \bibinfo {author} {\bibfnamefont {G.}~\bibnamefont {Tarjus}},\ }\bibfield  {title} {\bibinfo {title} {The ferromagnetic q-state potts model on three-dimensional lattices: a study for real values of q},\ }\href {https://doi.org/https://doi.org/10.1016/S0378-4371(01)00177-7} {\bibfield  {journal} {\bibinfo  {journal} {Physica A: Statistical Mechanics and its Applications}\ }\textbf {\bibinfo {volume} {296}},\ \bibinfo {pages} {460} (\bibinfo {year} {2001})}\BibitemShut {NoStop}%
\bibitem [{\citenamefont {Hartmann}(2005)}]{Hartmann2005}%
  \BibitemOpen
  \bibfield  {author} {\bibinfo {author} {\bibfnamefont {A.~K.}\ \bibnamefont {Hartmann}},\ }\bibfield  {title} {\bibinfo {title} {Calculation of partition functions by measuring component distributions},\ }\href {https://doi.org/10.1103/PhysRevLett.94.050601} {\bibfield  {journal} {\bibinfo  {journal} {Phys. Rev. Lett.}\ }\textbf {\bibinfo {volume} {94}},\ \bibinfo {pages} {050601} (\bibinfo {year} {2005})}\BibitemShut {NoStop}%
\bibitem [{\citenamefont {Gaite}(2024)}]{gaite2024}%
  \BibitemOpen
  \bibfield  {author} {\bibinfo {author} {\bibfnamefont {J.}~\bibnamefont {Gaite}},\ }\href {https://arxiv.org/abs/2407.15497} {\bibinfo {title} {New renormalization group study of the 3-state potts model and related statistical models}} (\bibinfo {year} {2024}),\ \Eprint {https://arxiv.org/abs/2407.15497} {arXiv:2407.15497 [cond-mat.stat-mech]} \BibitemShut {NoStop}%
\bibitem [{\citenamefont {Gavai}\ and\ \citenamefont {Karsch}(1992{\natexlab{a}})}]{Gavai1992}%
  \BibitemOpen
  \bibfield  {author} {\bibinfo {author} {\bibfnamefont {R.~V.}\ \bibnamefont {Gavai}}\ and\ \bibinfo {author} {\bibfnamefont {F.}~\bibnamefont {Karsch}},\ }\bibfield  {title} {\bibinfo {title} {Z(3) criticality in three dimensions: Study of extended potts models},\ }\href {https://doi.org/10.1103/PhysRevB.46.944} {\bibfield  {journal} {\bibinfo  {journal} {Phys. Rev. B}\ }\textbf {\bibinfo {volume} {46}},\ \bibinfo {pages} {944} (\bibinfo {year} {1992}{\natexlab{a}})}\BibitemShut {NoStop}%
\bibitem [{\citenamefont {{Binder}}(1981)}]{1981ZPhyB..43..119B}%
  \BibitemOpen
  \bibfield  {author} {\bibinfo {author} {\bibfnamefont {K.}~\bibnamefont {{Binder}}},\ }\bibfield  {title} {\bibinfo {title} {{Finite size scaling analysis of ising model block distribution functions}},\ }\href {https://doi.org/10.1007/BF01293604} {\bibfield  {journal} {\bibinfo  {journal} {Zeitschrift fur Physik B Condensed Matter}\ }\textbf {\bibinfo {volume} {43}},\ \bibinfo {pages} {119} (\bibinfo {year} {1981})}\BibitemShut {NoStop}%
\bibitem [{\citenamefont {Ma}\ \emph {et~al.}(2018)\citenamefont {Ma}, \citenamefont {Weinberg}, \citenamefont {Shao}, \citenamefont {Guo}, \citenamefont {Yao},\ and\ \citenamefont {Sandvik}}]{PhysRevLett.121.117202}%
  \BibitemOpen
  \bibfield  {author} {\bibinfo {author} {\bibfnamefont {N.}~\bibnamefont {Ma}}, \bibinfo {author} {\bibfnamefont {P.}~\bibnamefont {Weinberg}}, \bibinfo {author} {\bibfnamefont {H.}~\bibnamefont {Shao}}, \bibinfo {author} {\bibfnamefont {W.}~\bibnamefont {Guo}}, \bibinfo {author} {\bibfnamefont {D.-X.}\ \bibnamefont {Yao}},\ and\ \bibinfo {author} {\bibfnamefont {A.~W.}\ \bibnamefont {Sandvik}},\ }\bibfield  {title} {\bibinfo {title} {Anomalous quantum-critical scaling corrections in two-dimensional antiferromagnets},\ }\href {https://doi.org/10.1103/PhysRevLett.121.117202} {\bibfield  {journal} {\bibinfo  {journal} {Phys. Rev. Lett.}\ }\textbf {\bibinfo {volume} {121}},\ \bibinfo {pages} {117202} (\bibinfo {year} {2018})}\BibitemShut {NoStop}%
\bibitem [{\citenamefont {Shao}\ \emph {et~al.}(2016)\citenamefont {Shao}, \citenamefont {Guo},\ and\ \citenamefont {Sandvik}}]{doi:10.1126/science.aad5007}%
  \BibitemOpen
  \bibfield  {author} {\bibinfo {author} {\bibfnamefont {H.}~\bibnamefont {Shao}}, \bibinfo {author} {\bibfnamefont {W.}~\bibnamefont {Guo}},\ and\ \bibinfo {author} {\bibfnamefont {A.~W.}\ \bibnamefont {Sandvik}},\ }\bibfield  {title} {\bibinfo {title} {Quantum criticality with two length scales},\ }\href {https://doi.org/10.1126/science.aad5007} {\bibfield  {journal} {\bibinfo  {journal} {Science}\ }\textbf {\bibinfo {volume} {352}},\ \bibinfo {pages} {213} (\bibinfo {year} {2016})},\ \Eprint {https://arxiv.org/abs/https://www.science.org/doi/pdf/10.1126/science.aad5007} {https://www.science.org/doi/pdf/10.1126/science.aad5007} \BibitemShut {NoStop}%
\bibitem [{\citenamefont {Luck}(1985)}]{PhysRevB.31.3069}%
  \BibitemOpen
  \bibfield  {author} {\bibinfo {author} {\bibfnamefont {J.~M.}\ \bibnamefont {Luck}},\ }\bibfield  {title} {\bibinfo {title} {Corrections to finite-size-scaling laws and convergence of transfer-matrix methods},\ }\href {https://doi.org/10.1103/PhysRevB.31.3069} {\bibfield  {journal} {\bibinfo  {journal} {Phys. Rev. B}\ }\textbf {\bibinfo {volume} {31}},\ \bibinfo {pages} {3069} (\bibinfo {year} {1985})}\BibitemShut {NoStop}%
\bibitem [{\citenamefont {Iino}\ \emph {et~al.}(2019)\citenamefont {Iino}, \citenamefont {Morita}, \citenamefont {Kawashima},\ and\ \citenamefont {Sandvik}}]{Iino2019}%
  \BibitemOpen
  \bibfield  {author} {\bibinfo {author} {\bibfnamefont {S.}~\bibnamefont {Iino}}, \bibinfo {author} {\bibfnamefont {S.}~\bibnamefont {Morita}}, \bibinfo {author} {\bibfnamefont {N.}~\bibnamefont {Kawashima}},\ and\ \bibinfo {author} {\bibfnamefont {A.~W.}\ \bibnamefont {Sandvik}},\ }\bibfield  {title} {\bibinfo {title} {Detecting signals of weakly first-order phase transitions in two-dimensional potts models},\ }\href {https://doi.org/10.7566/jpsj.88.034006} {\bibfield  {journal} {\bibinfo  {journal} {Journal of the Physical Society of Japan}\ }\textbf {\bibinfo {volume} {88}},\ \bibinfo {pages} {034006} (\bibinfo {year} {2019})}\BibitemShut {NoStop}%
\bibitem [{\citenamefont {Cardy}(1984)}]{Cardy1984}%
  \BibitemOpen
  \bibfield  {author} {\bibinfo {author} {\bibfnamefont {J.~L.}\ \bibnamefont {Cardy}},\ }\bibfield  {title} {\bibinfo {title} {Conformal invariance and universality in finite-size scaling},\ }\href {https://doi.org/10.1088/0305-4470/17/7/003} {\bibfield  {journal} {\bibinfo  {journal} {Journal of Physics A: Mathematical and General}\ }\textbf {\bibinfo {volume} {17}},\ \bibinfo {pages} {L385} (\bibinfo {year} {1984})}\BibitemShut {NoStop}%
\bibitem [{\citenamefont {Cardy}(1985)}]{Cardy1985a}%
  \BibitemOpen
  \bibfield  {author} {\bibinfo {author} {\bibfnamefont {J.~L.}\ \bibnamefont {Cardy}},\ }\bibfield  {title} {\bibinfo {title} {Universal amplitudes in finite-size scaling: generalisation to arbitrary dimensionality},\ }\href {https://doi.org/10.1088/0305-4470/18/13/005} {\bibfield  {journal} {\bibinfo  {journal} {Journal of Physics A: Mathematical and General}\ }\textbf {\bibinfo {volume} {18}},\ \bibinfo {pages} {L757} (\bibinfo {year} {1985})}\BibitemShut {NoStop}%
\bibitem [{\citenamefont {Han}\ \emph {et~al.}(2023{\natexlab{b}})\citenamefont {Han}, \citenamefont {Hu}, \citenamefont {Zhu},\ and\ \citenamefont {He}}]{Four_Han2023}%
  \BibitemOpen
  \bibfield  {author} {\bibinfo {author} {\bibfnamefont {C.}~\bibnamefont {Han}}, \bibinfo {author} {\bibfnamefont {L.}~\bibnamefont {Hu}}, \bibinfo {author} {\bibfnamefont {W.}~\bibnamefont {Zhu}},\ and\ \bibinfo {author} {\bibfnamefont {Y.-C.}\ \bibnamefont {He}},\ }\bibfield  {title} {\bibinfo {title} {Conformal four-point correlators of the three-dimensional ising transition via the quantum fuzzy sphere},\ }\href {https://doi.org/10.1103/PhysRevB.108.235123} {\bibfield  {journal} {\bibinfo  {journal} {Phys. Rev. B}\ }\textbf {\bibinfo {volume} {108}},\ \bibinfo {pages} {235123} (\bibinfo {year} {2023}{\natexlab{b}})}\BibitemShut {NoStop}%
\bibitem [{\citenamefont {Delfino}\ and\ \citenamefont {Cardy}(2000)}]{Delfino2000}%
  \BibitemOpen
  \bibfield  {author} {\bibinfo {author} {\bibfnamefont {G.}~\bibnamefont {Delfino}}\ and\ \bibinfo {author} {\bibfnamefont {J.}~\bibnamefont {Cardy}},\ }\bibfield  {title} {\bibinfo {title} {The field theory of the q$\rightarrow$4+ potts model},\ }\href {https://doi.org/https://doi.org/10.1016/S0370-2693(00)00548-7} {\bibfield  {journal} {\bibinfo  {journal} {Physics Letters B}\ }\textbf {\bibinfo {volume} {483}},\ \bibinfo {pages} {303} (\bibinfo {year} {2000})}\BibitemShut {NoStop}%
\bibitem [{\citenamefont {Gavai}\ and\ \citenamefont {Karsch}(1992{\natexlab{b}})}]{PhysRevB.46.944}%
  \BibitemOpen
  \bibfield  {author} {\bibinfo {author} {\bibfnamefont {R.~V.}\ \bibnamefont {Gavai}}\ and\ \bibinfo {author} {\bibfnamefont {F.}~\bibnamefont {Karsch}},\ }\bibfield  {title} {\bibinfo {title} {Z(3) criticality in three dimensions: Study of extended potts models},\ }\href {https://doi.org/10.1103/PhysRevB.46.944} {\bibfield  {journal} {\bibinfo  {journal} {Phys. Rev. B}\ }\textbf {\bibinfo {volume} {46}},\ \bibinfo {pages} {944} (\bibinfo {year} {1992}{\natexlab{b}})}\BibitemShut {NoStop}%
\bibitem [{\citenamefont {Brown}(1989)}]{BROWN1989412}%
  \BibitemOpen
  \bibfield  {author} {\bibinfo {author} {\bibfnamefont {F.~R.}\ \bibnamefont {Brown}},\ }\bibfield  {title} {\bibinfo {title} {Absence of z3 criticality in a three-dimensional potts model with extended couplings},\ }\href {https://doi.org/https://doi.org/10.1016/0370-2693(89)91469-X} {\bibfield  {journal} {\bibinfo  {journal} {Physics Letters B}\ }\textbf {\bibinfo {volume} {224}},\ \bibinfo {pages} {412} (\bibinfo {year} {1989})}\BibitemShut {NoStop}%
\end{thebibliography}%

\end{document}